\documentclass[prb,a4paper,onecolumn,noshowpacs,citeautoscript,lineno,superscriptaddress]{revtex4-2}

\usepackage{amsmath,amssymb}
\usepackage{graphicx}
\usepackage{mathtools}
\usepackage{bm}

\renewcommand{\sb}[1]{_{\text {#1}}}  
\def\Sb#1{_{\scriptscriptstyle\rm{#1}}}

\def\nus{\nu_{\rm s}}
\def\vec#1{\bm{#1}}
\def\omk{\omega_{\vec k}}
\def\gak{\gamma_{\vec k}}
\def\KW{^{^{\rm KW}}}
\def\IW{^{^{\rm IW}}}
\def\omex{\omega_{\rm ex}}
\def\Oms{\Omega_{\rm s}}
\def\Omf{\Omega_{\rm f}}
\def\taus{\tau_{\rm s}}
\def\tf{t_{\rm g}^*}

\begin{document}

\title{Rotating quantum wave turbulence}

\author{J.T.~M\"akinen}
\email[]{jere.makinen@aalto.fi}
\affiliation{Department of Applied Physics, Aalto University, FI-00076 AALTO, Finland}

\author{S.~Autti}
\affiliation{Department of Applied Physics, Aalto University, FI-00076 AALTO, Finland}
\affiliation{Department of Physics, Lancaster University, Lancaster LA1 4YB, UK}

\author{P.J.~Heikkinen}
\affiliation{Department of Applied Physics, Aalto University, FI-00076 AALTO, Finland}
\affiliation{Department of Physics, Royal Holloway, University of London, Egham, Surrey, TW20 0EX, UK}

\author{J.J.~Hosio}
\affiliation{Department of Applied Physics, Aalto University, FI-00076 AALTO, Finland}

\author{R.~H\"anninen}
\affiliation{Department of Applied Physics, Aalto University, FI-00076 AALTO, Finland}
\affiliation{Finnish Meteorological Institute, P.O. BOX 503, FI-00101 Helsinki, Finland}

\author{V.S.~L'vov}
\affiliation{Department of Chemical Physics, Weizmann Institute of Science, Israel}

\author{P.M.~Walmsley}
\affiliation{School of Physics and Astronomy, The University of Manchester, UK}

\author{V.V.~Zavjalov}
\affiliation{Department of Applied Physics, Aalto University, FI-00076 AALTO, Finland}
\affiliation{Department of Physics, Lancaster University, Lancaster LA1 4YB, UK}

\author{V.B.~Eltsov}
\affiliation{Department of Applied Physics, Aalto University, FI-00076 AALTO, Finland}


\date{\today}

\maketitle


\textbf{
Rotating turbulence is ubiquitous in nature. Previous works suggest that such turbulence could be described as an ensemble of interacting inertial waves across a wide range of length scales. For turbulence in macroscopic quantum condensates, the nature of the transition between the quasiclassical dynamics at large scales and the corresponding dynamics at small scales, where the quantization of vorticity is essential, remains an outstanding unresolved question. Here we expand the paradigm of wave-driven turbulence to rotating quantum fluids where the spectrum of waves extends to microscopic scales as Kelvin waves on quantized vortices. We excite inertial waves at the largest scale by periodic modulation of the angular velocity and observe dissipation-independent transfer of energy to smaller scales and the eventual onset of the elusive Kelvin-wave cascade at the lowest temperatures. We further find that energy is pumped to the system through a boundary layer distinct from the classical Ekman layer and support our observations with numerical simulations. Our experiments demonstrate a new regime of turbulent motion in quantum fluids where the role of vortex reconnections can be neglected, thus stripping the transition between the classical and the quantum regimes of turbulence down to its bare bones.
}

\vspace{0.1cm}

\begin{figure} [tb]
\begin{center}
\includegraphics[width=0.6\linewidth]{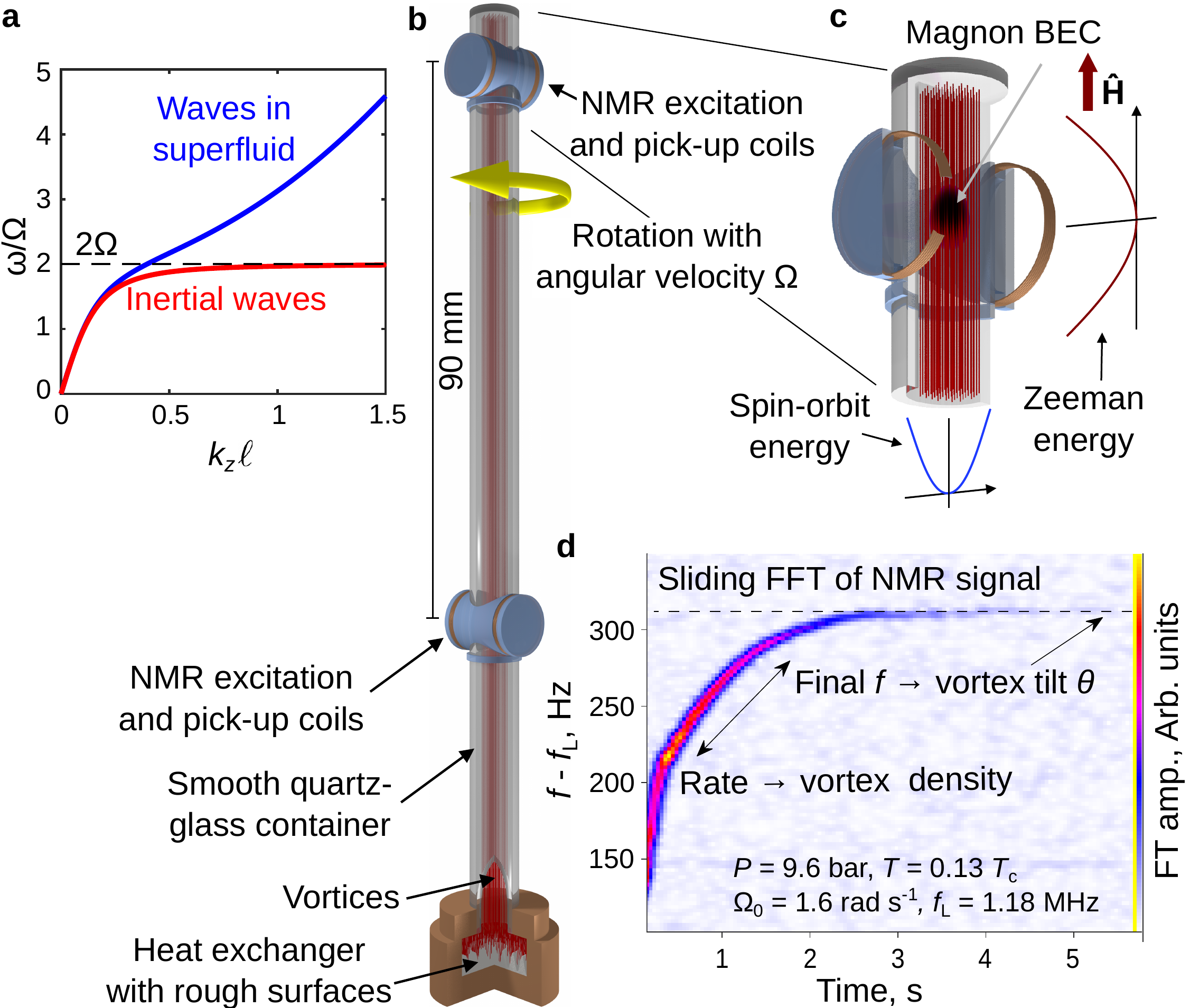} 
\end{center}
\caption{{\bf Experimental principles.} \textbf{a}, In superfluids, for fixed radial wave number, the full dispersion relation (blue line) extends beyond the classical IW regime (red line) with a cutoff frequency $2 \Omega$ (black dashed line) set by the angular velocity $\Omega$. Here $\omega$ is the angular frequency of the wave mode. \textbf{b}, A smooth-walled quartz-glass cylinder, filled with superfluid $^3$He-B, is rotated about its longitudinal axis. During the experiments we monitor the vortex configuration in two locations separated by 90$\,$mm using two pairs of NMR pick-up and excitation coils. The quartz glass container is open from the bottom to a heat exchanger volume with rough silver-sintered surfaces. \textbf{c}, The spatial distribution of vortices is monitored with a magnon Bose-Einstein condensate (BEC), trapped in the axial direction in a minimum of the magnetic field and in the radial direction by spatial variation of the spin-orbit energy (called texture). The radial trapping potential is modified by the presence of vortices. \textbf{d}, We use pulsed NMR to probe the ground state frequency in the magneto-textural trap. The frequency is shown as shift from the Larmor frequency $f_{\rm L}$. The relaxation rate of the signal depends on the vortex density \cite{PhysRevResearch.3.L032002}, while the final frequency is affected by the orientation of vortices (see SM for details).}
\label{fig:setup}
\end{figure}

\begin{figure*} [!ht]
\centerline{\includegraphics[width=\linewidth]{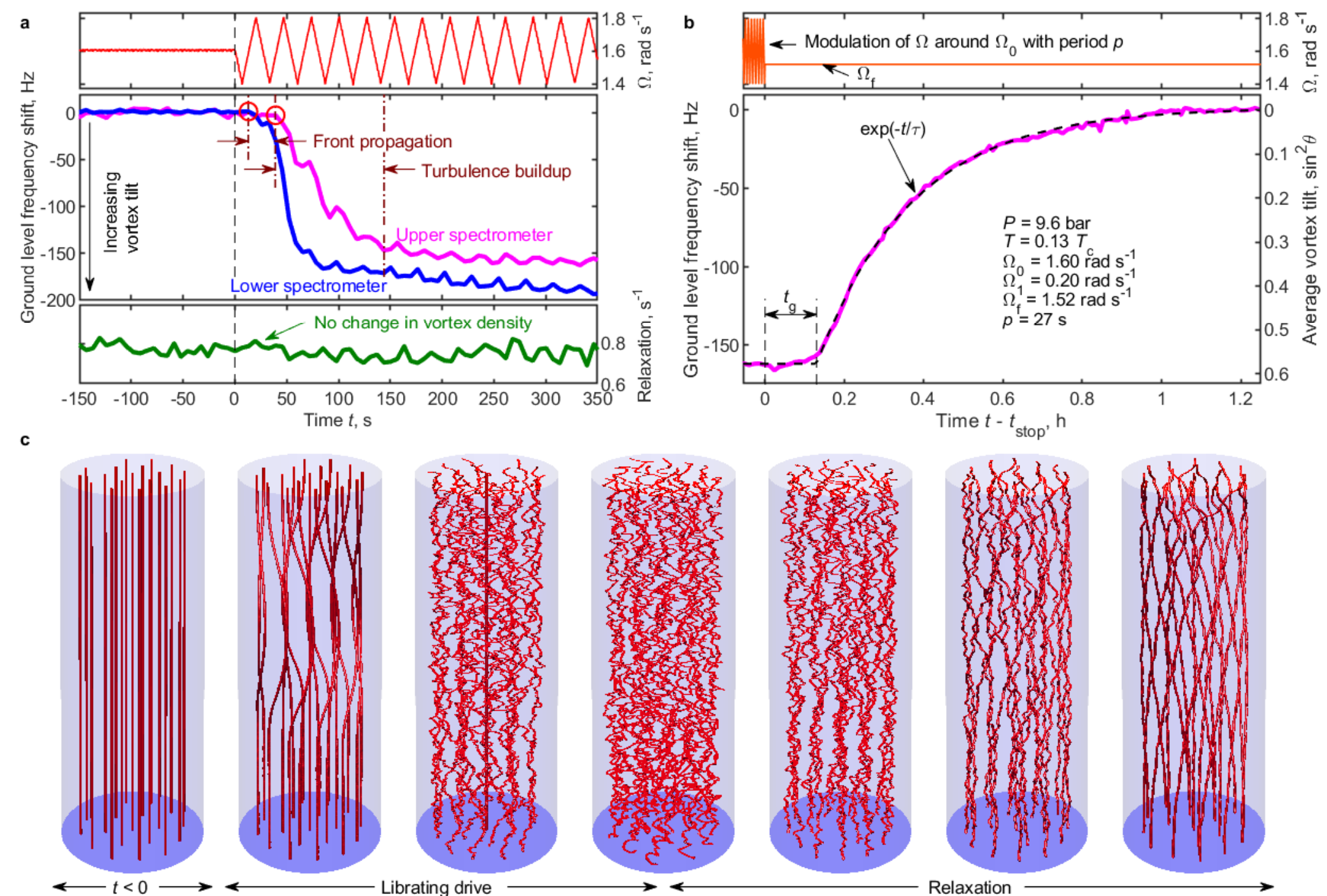}}
\caption{{\bf Buildup and decay of rotating quantum wave turbulence.} \textbf{a}, Modulation of the angular velocity (top panel) results in a propagating front of increasing vortex tilt, seen as a frequency shift in NMR measurements (middle panel). The front velocity and saturation time of the signal are explained by dynamics of IWs (see SM). The NMR relaxation rate in the lower spectrometer (bottom panel) shows no observable change in response to the librating drive, indicating that the vortex density remains constant during this time \cite{PhysRevResearch.3.L032002}. \textbf{b}, At time $t_{\rm stop}$ the drive is ramped to a final angular velocity $\Omega_{\mathrm{f}}$ using the same acceleration $\dot{\Omega}$ as during the modulations. After that an initial period of duration $t_{\rm g}$, during which the average vortex tilt angle $\theta$ remains at a level comparable to the developed turbulence, is followed by an exponential decay towards the equilibrium state. Our estimate (see SM for details) for the mean vortex tilt angle within the upper spectrometer in units of $\sin^2\theta$ is shown on the right axis. Panels {\bf a} and {\bf b} correspond to the same experimental run with $t_{\rm stop} \approx 1.0\,$h. {\bf c}, We qualitatively confirm our interpretations in vortex filament simulations, where vortex lines (red) in a cylindrical container (light blue, $L = 50\,$mm, $\ell \sim 0.1\,$mm), are driven out of equilibrium by applying a rotational drive similar to that in the experiments (figures not to scale). The drive couples to the vortices via a thin layer with high mutual friction at the bottom (dark blue). The horizontal alignment of the figures correspond to the state of the experiment right above it.}
\label{fig:startstopmod}
\end{figure*}

\begin{figure*} [!t]
\centerline{\includegraphics[width=1\linewidth]{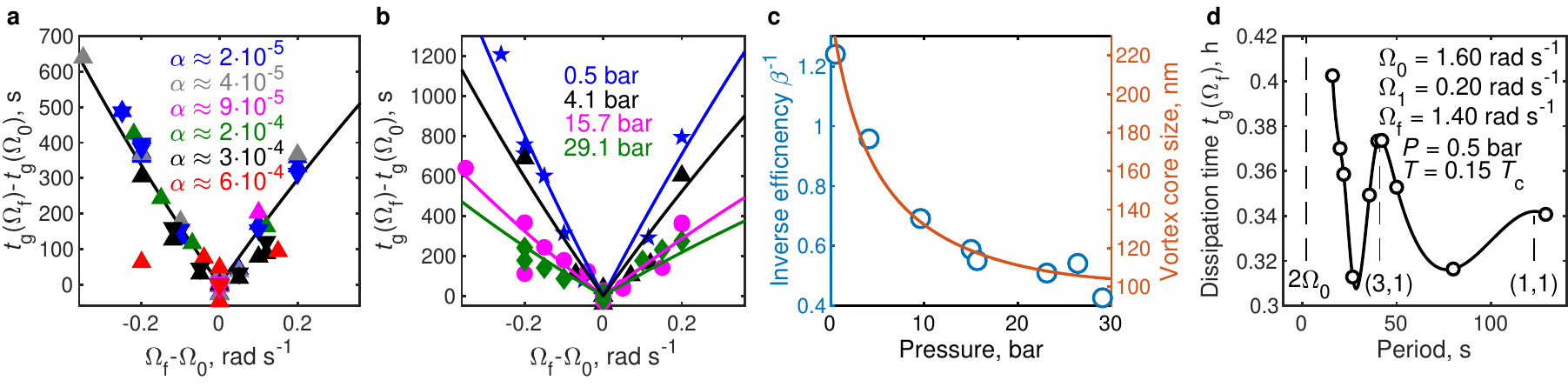}}
\caption{{\bf Transfer of energy through the quantum boundary layer.} \textbf{a}, Observed relaxation times of the energy stored in solid-body-like rotating flow (symbols) are compared with the model of pumping by the quantum boundary layer characterized by the time $t_{\rm g}^*$ (solid black line). Here $\Omega_0 = 1.6\,$rad$\,$s$^{-1}$, $\Omega_1 = 0.20\,$rad$\,$s$^{-1}$, $p = 27\,$s, the plot contains data measured in both $15.0\,$bar and $15.7\,$bar pressure, and the fitted energy pumping efficiency is $\beta^{-1} = 0.57$. The difference between the time $t_{\rm g}(\Omega_{\rm f})$ and the value $t_{\rm g}(\Omega_0)$ separately measured with $\Omega_{\rm f}=\Omega_0$ is plotted, to account for any excess energy (such as near-resonant inertial waves or geostrophic modes) not present in our model. Different colors correspond to different temperatures between $(0.14-0.19)\,T_{\rm c}$ and thus, values of the mutual friction parameter $\alpha$, marked in the figure. Upwards-pointing triangles correspond to data measured with the upper spectrometer, while downwards-pointing triangles correspond to data from the lower spectrometer. \textbf{b}, The pumping efficiency is found to change with pressure. \textbf{c}, The observed increase of $\beta$ with pressure (symbols, left axis) may be understood via enhanced pinning with decreasing vortex core size \cite{PhysRevLett.115.235301} (solid line, right axis). \textbf{d}, Measurements of $t_{\rm g}$ with different period of excitation show an increase in the stored energy in the presence of standing inertial wave modes with low axial wave numbers. The IW cutoff frequency and the corresponding IW modes ($M,N$), where $M$ is the axial and $N$ is the radial wave number, are marked with vertical dashed lines.}
\label{t0-vels}
\end{figure*}

Rotating turbulence plays an important role in systems such as planets' atmospheres \cite{HOEKSTRA19992055, RedSpot_exp, RedSpot_simu}, turbomachinery \cite{BRADSHAW1996575}, rotating quantum gases \cite{kobuszewski2020rotating}, and neutron stars \cite{NeutronStarTurb, NeutronStarTurb2}. Generally speaking, rotating flows of incompressible classical fluids can be characterized by two dimensionless numbers: the Reynolds number $\mathrm{Re}$ denoting the ratio of inertial to dissipative forces, and the Rossby number $\mathrm{Ro}$ expressing the ratio of inertial forces to the Coriolis force. In the limit ${\rm Re} \gg 1$ the flow becomes turbulent, while for ${\rm Ro} \ll 1$ the Coriolis force becomes dominant over the inertial forces and therefore the rotational effects are strong. In superfluids the Rossby number can be defined in a similar fashion, while the physical meaning of the Reynolds number is captured by the 'superfluid Reynolds number' \cite{superfluid_reynolds} $\mathrm{Re}_{\alpha}$ which only depends on {\it intrinsic} mutual friction parameters that describe the coupling between the quantized vortices and the normal component \cite{PhysRevB.97.014527}. Theoretical \cite{PhysRevE.68.015301}, numerical \cite{bellet_godeferd_scott_cambon_2006}, and experimental \cite{iwturb_nphys,PhysRevFluids.2.122601,PhysRevLett.125.254502,PhysRevLett.124.124501} work suggests that rotating turbulence, for which $\mathrm{Re} \gg 1$ and $\mathrm{Ro} \ll 1$, could be described as an ensemble of interacting inertial waves (IW) in classical fluids. The measurements presented here cover $\mathrm{Ro} \sim (1-3) \cdot 10^{-2}$ and $\mathrm{Re}_{\alpha} \sim 10^3 - 10^5$, which puts our experimental conditions well within the inertial wave turbulence regime.

Quantum turbulence is usually considered as a complex dynamic tangle of reconnecting quantized vortices \cite{PhysRevLett.118.134501,PhysRevLett.100.245301,PhysRevA.96.023617,Barenghi4647}. In the regime $\mathrm{Re}_{\alpha} \gg 1$ and $\mathrm{Ro} \ll 1$ quantized vortices are nearly parallel and inter-vortex reconnections are suppressed \cite{PhysRevB.76.024520}, exposing the underlying wave-turbulent energy cascade to experimental observation. At the largest length scales the superfluid flow field may mimic that of classical IWs via collective motion of quantized vortices. Contrary to classical fluids, in superfluids the spectrum of waves extends beyond the IW cutoff frequency, Fig.~\ref{fig:setup}{\bf a}, as Kelvin waves (KW) \cite{kelvin,hendersonbarenghi} carried by individual vortices. The crossover between these regimes takes place at $k_{z} \ell \sim 0.5$, where $k_z$ is the axial wave vector and $\ell$ is the mean inter-vortex distance set by the angular velocity.

To highlight differences between classical and quantum turbulence, the low-temperature limit is of particular interest since negligible frictional forces allow transfer of energy to length scales where quantization of vorticity is essential \cite{Skrbeke2018406118,Barenghi4647}. In this limit, the energy is believed to flow towards the smallest scales through a cascade of KWs \cite{PhysRevLett.92.035301, Lvov2010} or through a quantum stress cascade \cite{PhysRevE.103.023106} and is ultimately dissipated via emission of sound waves \cite{Vinen2005}, emission of quasiparticles \cite{PhysRevLett.108.045303,PhysRevB.44.9667}, or, at a finite temperature, via mutual friction \cite{PhysRevB.97.014527, Bou__2012}. Despite observations of vortex reconnections and the related production of KWs \cite{Bewley13707,Fonda4707}, direct experimental proof of the existence of the KW cascade has remained elusive.

In the experiments we initially rotate the sample volume, Fig.~\ref{fig:setup}\textbf{b}, with a constant angular velocity to create an array of quantized vortices with aerial density $\ell^{-2}$ oriented along the axis of rotation. We monitor the vortex configuration independently at two spatially separated locations, Fig.~\ref{fig:setup}\textbf{c}, via pulsed nuclear magnetic resonance (NMR) techniques, Fig.~\ref{fig:setup}\textbf{d}. We then perturb the vortex array by applying a librating drive $\Omega(t) = \Omega_{0} + \Omega_{1} f(\omega_{\mathrm{ex}} t)$, where $\Omega_{0}$ is the mean angular velocity during the drive with amplitude $\Omega_1 < \Omega_0$, and $f$ describes a triangle wave between $[-1,1]$ with period $p=2 \pi \omega_{\mathrm{ex}}^{-1}$. During the librating drive, the following forces are exerted on the vortices: the force due to mutual friction, the Magnus force, and the force due to pinning of the vortex ends at the rough bottom of the container. Elsewhere the smooth walls of the cylinder allow nearly frictionless vortex sliding \cite{Hosio2012}.

Soon after we start the librating drive we observe an increase in the average vortex tilt angle $\theta$ with respect to the axis of rotation. Notably, a propagating wave front originates from the bottom of the container, Fig.~\ref{fig:startstopmod}\textbf{a}, with phase velocity $V\sb{prop}\approx 0.3\,$cm~s$^{-1}$. This velocity agrees with the phase velocity of the first axially symmetric radial inertial wave mode, $V\sb{ph} = \omega_{\rm ex}/k_z \approx 0.25\,$cm s$^{-1}$. Simultaneously, we observe no change in the relaxation of the NMR signal, indicating that the vortex density remains constant during this time \cite{PhysRevResearch.3.L032002}.

During the drive the action of hydrodynamic forces on a vortex would far exceed the maximum pinning force, equal to vortex tension $T_{\rm v} \sim 10^{-8}\,$cm$\cdot$g$\cdot$s$^{-2}$. In this case vortices are inevitably stretched and the rotating superfluid forms a {\it quantum boundary layer}, previously discussed in Ref.~\citenum{PhysRevB.32.171}, in which each vortex is acted upon with force equal in magnitude to $T_{\rm v}$. We describe the flow of energy in such a system using a phenomenological model (see SM for details) in which the quantum boundary layer pumps energy to a cascade of IWs \cite{ALEXAKIS20181}, which in turn feeds a cascade of KWs. In the KW regime, the energy is consumed by mutual friction, which also terminates the KW cascade. A qualitatively similar picture is obtained in vortex-filament calculations in the presence of a surface layer with increased mutual friction, Fig.~\ref{fig:startstopmod}{\bf c}. The calculations demonstrate how, in response to the drive, vortices, initially excited at the long wave length limit, transfer energy towards smaller scales while the role of vortex reconnections is negligible.

After the system has reached a steady state, we stop the librating drive and ramp the angular velocity to a chosen value $\Omf$. The vortex array then relaxes from its steady-state configuration with $\theta \sim 50^{\circ}$ towards the equilibrium state with $\theta\to 0$  via a process comprising two clearly distinct stages, Fig.~\ref{fig:startstopmod}\textbf{b}. In the first stage, $\theta$ remains at a similar level as during the drive. During this interval, marked as $t_{\rm g}$, large vortex tilt is sustained by feeding the global flow energy via the quantum boundary layer to IWs (and subsequently to KWs via the IW cascade). The second relaxation stage, an exponential restoration of the equilibrium vortex configuration, $\sin^2\theta \propto \exp(-t/\tau)$, takes place after the energy from the global flow has been consumed and the energy stored in the KW cascade is dissipated by mutual friction. This behavior is qualitatively reproduced in our model calculations (see SM).

The duration of the first relaxation stage is set by the amount of energy stored in the solid-body-like flow and controlled experimentally by varying the final angular velocity $\Omega_{\mathrm{f}}$. The characteristic time $t_{\rm g}(\Omega_{\rm f})$ increases with increasing $|\Omf-\Omega_0|$, Fig.~\ref{t0-vels}\textbf{a}. If each vortex is acted upon in a boundary layer with a force equal to $\beta T_{\nu}$ in magnitude, the superfluid angular velocity reaches $\Omf$ in a finite time $\tf = \beta^{-1} \taus  \left| \ln (\Omf/\Omega_0) \right|,$ where $\taus \approx 4\cdot10^3\,{\rm s}$ (see SM for the derivation) and fitting parameter $\beta \sim 1$ characterizes the energy pumping efficiency. The expression for $\tf$ agrees with our observations of $t_{\rm g}$ with a single fitted $\beta$ in the temperature range $(0.13\,$--$\,0.19)\,T_{\rm c}$, where $T_{\rm c} \sim 1\,$mK is the superfluid transition temperature of $^3$He. Over this temperature range the dissipative mutual friction parameter $\alpha$, which controls the energy dissipation rate in the bulk, changes by almost two orders of magnitude \cite{PhysRevB.97.014527}. This temperature independence confirms our picture of the quantum boundary layer feeding the IW energy cascade. As a function of pressure, the observed change in $\beta$, Figs.~\ref{t0-vels}\textbf{b}-\textbf{c}, could be explained by the change in the vortex core size \cite{PhysRevLett.115.235301} with the premise that smaller core size results in enhanced pinning. In addition, the complicated dependence of $t_{\rm g}$ on $\omex$, Fig.~\ref{t0-vels}\textbf{d}, may be understood as additional contributions to the energy of the global flow in the vicinity of standing axially symmetric inertial wave resonances in the cylindrical sample container \cite{fultz1959, PhysRevB.86.060518} and possible generation of geostrophic modes \cite{PhysRevLett.124.124501,PhysRevLett.125.254502}.

\begin{figure*} [t]
\centerline{\includegraphics[width=1\linewidth]{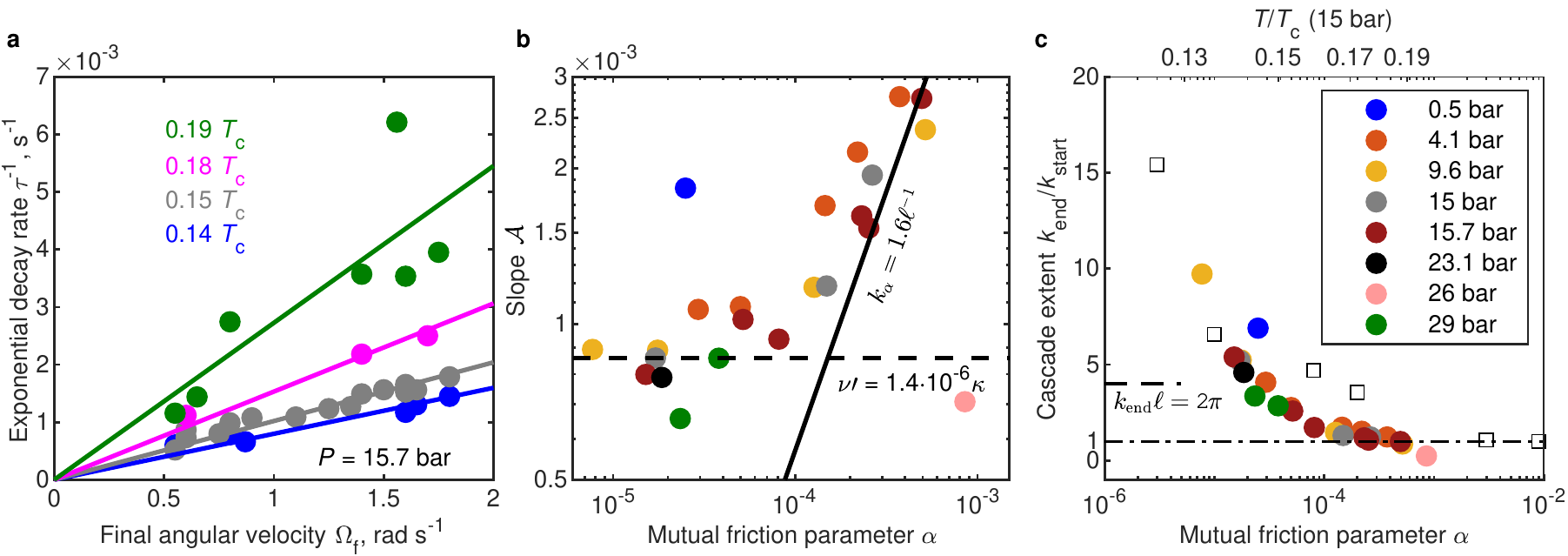}}
\caption{{\bf Formation of the KW cascade with decreasing temperature.} \textbf{a}, The observed exponential relaxation time constant scales as $\tau^{-1} = \mathcal{A} \Omega_{\mathrm{f}}$. The relaxation is faster at higher temperatures, in contrast to the temperature-independent duration $t_{\rm g}$ of the initial relaxation stage in Fig.~\ref{t0-vels}\textbf{a}. For these measurements, we kept $\Omega_1 = 0.2\,$rad$\,$s$^{-1}$ and $\Omega_0\dot{\Omega}^{-1} \approx 53\,$s constant. \textbf{b}, The mutual-friction dependence of the relaxation rate normalized to $\Omf$, that is, the slope $\mathcal{A}$ of lines in the \textbf{a} panel, deviates from the linear dependence obtained by fixing the dissipativative length scale by $k_\alpha = 1.6\ell^{-1}$ (solid line). This deviation suggests that $k_{\alpha}$ is changing with temperature as a result of the KW cascade. From data at the lowest temperatures, we estimate the value of the effective kinematic viscosity $\nu' \sim 1.4 \cdot 10^{-6} \kappa$, marked with the dashed line (see SM for details). \textbf{c}, We extract the extent of the KW cascade (filled circles) in the $k$ space as discussed in the main text. We observe the extension of the cascade further in the $k$ space with decreasing $\alpha$. Our observations are compared to numerical simulations (empty squares) from Ref.~\citenum{PhysRevB.91.144501}. At the lowest temperatures ($\alpha \lesssim 3 \cdot 10^{-5}$), the KW cascade extends to the length scales smaller than the inter-vortex distance (shown as the short dashed line). Symbol colors in panels {\bf b} and {\bf c} mark different pressures shown in the legend in the panel {\bf c}.}
\label{small_scales}
\end{figure*}

Let us now turn our attention to the second relaxation stage. For a single KW with a wave vector $k_{\alpha}$, the energy dissipation rate by mutual friction is exponential with the decay rate \cite{donnelly} $\tau_\alpha^{-1} = 2 \alpha \nu_{\rm s} k_{\alpha}^2$, where $\nus \sim\,4\cdot 10^{-4}\,$cm$^2$s$^{-1}$. On the other hand, for a distribution of KWs in the form of a cascade extending between $k_{\rm start}$ and $k_{\rm end} \gg k_{\rm start}$, the dissipation remains exponential with the rate given by $\tau_\alpha^{-1}$ at an effective length scale $k_{\alpha} = k_{\rm start}^{1/3} k_{\rm end}^{2/3}$ (see SM for derivation). To distinguish between the two scenarios we study the dependence of $\tau_\alpha^{-1}$ on rotation velocity and temperature. We find that the relaxation rate is linearly proportional to $\Omf$ at a constant temperature (constant $\alpha$), i.e. $\tau^{-1} \equiv \mathcal{A} \Omf$, suggesting $k_\alpha \propto \ell^{-1}$, Fig.~\ref{small_scales}{\bf a}. Therefore, for a fixed $k_\alpha \ell$ (dissipative length scale set by $\ell$ in the absence of a cascade), $\mathcal{A}$ is expected to scale linearly in $\alpha$. We find that at higher values of $\alpha$ (higher temperatures), $\mathcal{A}$ is roughly linear in $\alpha$, Fig.~\ref{small_scales}{\bf b}. However, the deviation from the linear dependence towards the lowest $\alpha$ (lowest temperatures) implies that $k_\alpha$ changes with temperature. In the KW cascade picture this is naturally explained by extension of the cascade towards larger $k_{\rm end}$ with decreasing $\alpha$ \cite{Bou__2012}. Using values of $\alpha$ from Ref.~\citenum{PhysRevB.97.014527} (with $\alpha=0$ at $T=0$) we convert the measured $\mathcal{A}$ to effective $k_\alpha$ and then, assuming $k_{\rm start}$ equals $k_\alpha = 1.6\ell^{-1}$ found at higher temperatures, we obtain the extent of the KW cascade $k_{\mathrm{end}}/ k_{\mathrm{start}}$, Fig.~\ref{small_scales}\textbf{c}. The cascade quickly expands to larger wave vectors for $\alpha \lesssim 10^{-4}$, in agreement with previous numerical simulations \cite{PhysRevB.91.144501}.

Our experimental observations, namely that $k_\alpha \propto \ell ^{-1}$ and that $\tau^{-1}$ tends towards a constant value at the lowest $\alpha$, are consistent with theoretical predictions \cite{PhysRevB.76.024520} (see also SM) linking the extent of the KW cascade to the effective kinematic viscosity $\nu'$, used to characterize the energy dissipation rate in quantum turbulence. Using the lowest-temperature data in Fig.~\ref{small_scales}{\bf b}, we obtain an estimate $\nu' \sim 10^{-6} \kappa$, where $\kappa \approx 6.6\cdot 10^{-4}\,$cm$^2\,$s$^{-1}$ is the quantum of circulation in $^3$He. The obtained value is ﬁve orders of magnitude smaller than for homogeneous and isotropic quantum turbulence \cite{PhysRevB.94.094502}, highlighting the different nature of the turbulent flows. Smaller values of $\nu'$ are generally thought to originate from nearly parallel arrangement of vortices \cite{PhysRevB.76.024520} and in the absence of vortex reconnections \cite{PhysRevA.67.015601}, both of which are realized in our experiments. We also note that while a recent theoretical work \cite{PhysRevE.103.023106} put forward an idea of a 'quantum stress cascade' as a possible energy transfer mechanism, our observations -- in particular the magnitude of the average vortex tilt $\theta$ determined mostly by KWs, the temperature dependence of the dissipative length scale $k_\alpha$, and the wave vector range of the excited KWs from $k_{\rm start}$ to $k_{\rm end}$ -- imply the picture involving a cascade of KWs \cite{Lvov2010,PhysRevLett.92.035301}. However, the outliers in the higher-temperature data in Figs.~\ref{t0-vels} and \ref{small_scales} may indicate that this picture changes with increasing temperature for $\alpha \gtrsim 10^{-3}$ where the dissipative length scale, set by mutual friction, crosses over from quantum ($\lesssim \ell$) to classical ($\gtrsim \ell$) length scales and the KW cascade is completely suppressed.

In historical context our work relates to the centuries-old d'Alembert's paradox stating that for incompressible potential flow (applicable also to a superfluid) there is no drag for a body moving with constant velocity within the fluid. The solution to this apparent paradox was introduced by Prandtl, who noted that the coupling between the moving body and the surrounding fluid originates in thin surface layers. For rotating flows in classical systems such as the atmosphere or oceans, as a result of a 'no-slip' boundary condition, the surface layer takes the form of an Ekman layer. Our experimental findings in the presence of a rough surface are consistent with a 'partial-slip' quantum boundary layer, unique to superfluids, where the magnitude of the applied force per quantized vortex is limited to a constant value. On the other hand, superfluids in the zero temperature limit may allow for experimental realization of the original d'Alembert's paradox in the presence of a smooth surface \cite{SmoothSurfNote} or if vortices are immobilized in the whole volume e.g. by a nano-structured confinement \cite{PhysRevResearch.2.033013}. In this work, the presence of the quantum boundary layer allows us to excite vortex waves, which develop into a novel type of quantum turbulence driven by non-linear interactions between vortex waves instead of vortex reconnections. Finally, the measurements presented in Fig.~\ref{small_scales}\textbf{b} support the existence of the dissipative anomaly for a cascade of KWs. The dissipative anomaly is also referred to as the zeroth law of turbulence due to its fundamental importance for the turbulence theory, and it states that dissipation should remain finite even in the limit of vanishing viscosity (or infinite Reynolds number). However, its nature and very existence for various forms of turbulence is still an active topic of research \cite{sreenivasan_2021}.

\section*{Acknowledgements}

This work has been supported by the European Research Council (ERC) under the European Union's Horizon 2020 research and innovation programme (Grant Agreement No. 694248) and by Academy of Finland project No. 332964. Additionally, the research leading to these results has received funding from the European Union's Horizon 2020 research and innovation programme under Grant Agreement No. 824109. S.A. acknowledges support from the Jenny and Antti Wihuri Foundation via the Council of Finnish Foundations. V.S.L. was in part supported by NSF-BSF grant No. 2020765. The experiments were performed at the Low Temperature Laboratory, which is a part of the OtaNano research infrastructure of Aalto University and of the European Microkelvin Platform.

\section*{Author contributions}

The experiments were designed by J. T. M., J. J. H., P. M. W., and V. B. E.; the experiments were conducted by J. T. M., S. A., P. J. H., J. J. H., P. M. W., and V. V. Z.; the theoretical analysis was carried out by J. T. M., V. S. L., P. M. W., and V. B. E.; numerical calculations were performed by J. T. M. and R. H.; V. B. E. supervised the project; and the paper was written by J. T. M., V. S. L., P. M. W., and V. B. E., with contributions from all authors.






\makeatletter
\renewcommand*\@biblabel[1]{[M#1]}
\makeatother

\renewcommand{\citenumfont}[1]{M#1}

\section*{Methods}

\subsection*{Sample geometry and thermometry}

Our choice of liquid is the superfluid B-phase of $^3$He, which can be studied with non-invasive NMR methods and for which the low-temperature limit is experimentally accessible. The sample is confined within a 150-mm-long cylindrical container with $\varnothing$5.85~mm inner diameter, made from quartz glass, Fig.~\ref{fig:setup}~{\bf b-c}. To avoid vortex pinning on the walls of the container, its inner surfaces are treated with hydrofluoric acid \cite{Spierings1993M}. The experimental volume, filled with $^3$He-B, is open from the bottom for thermal coupling to the nuclear demagnetization stage. The experimental volume contains two commercial quartz tuning forks with $32$~kHz resonance frequency, commonly used for thermometry in $^3$He experiments \cite{Blaauwgeers2007M,Riekki2019M}. The forks are calibrated against the Leggett frequency of $^3$He-B, found by continuous wave NMR spectroscopy at $0.37 T_{\mathrm{c}}$ at 0.5 bar. At lower temperatures we assume that the forks' behavior is limited to the ballistic regime of quasiparticle propagation, where the forks' resonance width behaves as $\Delta f = \mathcal{C} \exp (-\Delta (k_{\mathrm{B}}T)^{-1}) + \Delta f_0$. Parameter $\mathcal{C} \sim 10 \pm 1.5$~kHz is the geometric factor and $\Delta f_{0} \sim 10-100$~mHz, determined by comparison to magnetic relaxation of the magnon BEC \cite{Heikkinen2014M}, is the forks' intrinsic width. The calibration is extrapolated to other pressures assuming $\mathcal{C} \propto p_{\mathrm{F}}^4$ \cite{Bradley2009M}, where $p_{\mathrm{F}}$ is the Fermi momentum.

\subsection*{NMR spectroscopy}

Vortex lines affect the spatial order-parameter distribution (texture) in superfluid $^3$He-B owing to contributions from the vortex cores and superflow around them \cite{Eltsov2011M}. Information about the order-parameter texture can be extracted via magnetic quasiparticles, magnons, pumped to a three-dimensional trapping potential with a radiofrequency pulse. The magnons quickly form a uniformly precessing Bose-Einstein condensate in the trap formed by the order-parameter texture in the radial direction and by a minimum of the magnetic field in the axial direction. The amplitude of the NMR signal is proportional to the number on magnons in the trap, which also affects the frequency of the signal. In rotation and at low temperatures, the lifetime of magnons in the trap is limited by conversion to other spin-wave modes mediated by vortices. Thus the decay time of the NMR signal is a measure of the vortex line density \cite{PhysRevResearch.3.L032002M}. Simultaneously, vortex orientation affects the textural part of the magnon trap and the energy of the ground state in the trap, which modifies the precession frequency of the magnon BEC seen in NMR.

In the measurements we use static magnetic field of 25$\,$mT  and 36$\,$mT in the upper and lower spectrometers, respectively. The corresponding NMR frequencies are 830$\,$kHz and 1.2~MHz. The magnetic field is created using coils whose symmetry axis is aligned along the axis of rotation. The NMR pick-up coils are spatially separated along the axis of rotation by 90 mm, oriented perpendicular to one another, and perpendicular to the axis of rotation. The upper pick-up coil is made of copper wire and is a part of the tank circuit with quality factor $Q \sim 1.5\cdot10^{2}$. The lower pick-up coil is made of superconducting wire and is a part of the tank circuit with quality factor $Q \sim 7.5\cdot 10^{3}$. We use cold preamplifiers, thermalized to a bath of liquid helium, and room temperature preamplifiers, to improve the signal-to-noise ratio in the measurements.

\subsection*{Rotating refrigerator}

The sample can be rotated about its vertical axis with angular velocities up to $3$~rad s$^{-1}$, and cooled down to $\sim 150 \mu$K using ROTA nuclear demagnetization refrigerator. The refrigerator is well balanced and suspended against vibrational noise. The earth's magnetic field is compensated using two saddle-shaped coils installed around the refrigerator to avoid parasitic heating of the nuclear stage. In rotation, the total heat leak to the sample remains below 20 pW \cite{PhysRevB.84.224501M}. The rotation velocity is typically changed with the rate $|\dot{\Omega}| = 0.03$~rad~s$^{-2}$.\\

\subsection*{Vortex filament simulations}

Vortex filament simulations \cite{PhysRevB.31.5782M, Hanninen4667M}, based on the Biot-Savart law, are used to support our interpretation of the experimental observations. The simulations start with 19 vortices initially distributed in three rings with one, six, and twelve vortices, from innermost to outermost ring, respectively. Initially the vortices are straight and terminate in the top and bottom walls, spanning a total of 50 mm each with spatial resolution of 0.125$\,$mm. The initial separation of the straight vortices corresponded to a rotating drive of $1.60\,$rad$\,$s$^{-1}$. An external periodic drive between $1.40$ and $1.80\,$rad~s$^{-1}$ with acceleration of $0.03$~rad~s$^{-2}$ is used to drive the vortices out of equilibrium. At the top and bottom boundaries, image vortices are used to prevent flow through the boundary. The vortices couple to the external drive via mutual friction ($\alpha = 1.77 \times 10^{-3}$ in the bulk and $\alpha = 2$ within a 0.1~mm layer at the bottom - highlighted with dark blue in Fig.~\ref{fig:startstopmod}{\bf d}). The high-friction boundary layer approximates vortex pinning at the bottom boundary in the experiments. In simulations, we observe an upwards-propagating wave similar to the experiments and the eventual development of vortex waves at small scales.

We note that during the drive (which was on for 600$\,$s) in the simulations there are a total of 127 inter-vortex reconnection events ($40\,\mu$m used as the reconnection distance), with an average reconnection rate $\sim 2 \cdot 10^{-3}\,$(cm$\cdot$s)$^{-1}$. Averaging over 20$\,$s intervals, the highest reconnection rate per vortex length is $\sim 10^{-2}\,$(cm$\cdot$s)$^{-1}$ or about 1 reconnection every 20 seconds per vortex. In addition, small scale structures appear before the first reconnection event takes place, suggesting that inter-vortex reconnections do not play a significant role on the development of the cascade. When the modulation of the rotation velocity is stopped (here $\Omf = \Omega_0$), the vortex configuration decays towards the equilibrium state with parallel straight vortices. To reduce the computation time for illustrative purposes, the two rightmost images in Fig.~\ref{fig:startstopmod}{\bf d} were obtained by developing the state with higher mutual friction ($\alpha = 4.7 \cdot 10^{-2}$). In the simulations we used core size $a_0 \approx 1.7\cdot 10^{-5}\,$mm.

\subsection*{Validity of weak turbulence theory}

A cascade of KWs can be described within the framework of weak turbulence theory (WTT) \cite{PhysRevE.68.015301M, nazarenko_waveturbM}, whose validity has been recently confirmed in experiments \cite{PhysRevLett.125.254502M}. In principle, WTT can be applied to a variety of systems, both classical \cite{iwturb_nphysM, PhysRevFluids.2.122601M, PhysRevLett.125.254502M, PhysRevLett.124.124501M, Brazhnikov_2002M, PhysRevLett.107.214503M,PhysRevLett.103.044501M} and superfluid \cite{Efimov2000M,PhysRevLett.76.2105M, KOLMAKOV1995470M, Navon2016M, Kolmakov4727M}, given that proper experimental conditions are met. Rotating quantum wave turbulence differs from hydrodynamic quantum turbulence \cite{PhysRevLett.118.134501M,PhysRevLett.100.245301M,PhysRevA.96.023617M,Barenghi4647M} in that the energy cascade is driven by nonlinear interactions between waves instead of vortex reconnections. We note that while in our experiments the average vortex tilt angle during the steady state is relatively large, $\theta \sim 50^{\circ}$, and thus we do not expect WTT to hold initially, the second relaxation stage in Fig.~\ref{fig:startstopmod}\textbf{b} nonetheless proceeds throughout with a single exponent. When only the tail of the relaxation (with $\theta < 35^\circ$ and thus closer to weak turbulence regime) is processed, only increase in scatter without qualitative changes is observed. We therefore fit the whole decay for the purposes of extracting the decay time constant $\tau$.




\newpage

\makeatletter
\renewcommand*\@biblabel[1]{[S#1]}
\makeatother

\renewcommand{\citenumfont}[1]{S#1}

\renewcommand{\figurename}{\textbf{Supplementary Figure}}
\setcounter{figure}{0}
\makeatletter 
\renewcommand{\thefigure}{S\@arabic\c@figure}
\makeatother

\renewcommand\theequation{S\arabic{equation}}

\section*{Supplemental Material}

\subsection*{Interpretation of the NMR measurement}

\begin{figure} [tb]
\includegraphics[width=0.45\linewidth]{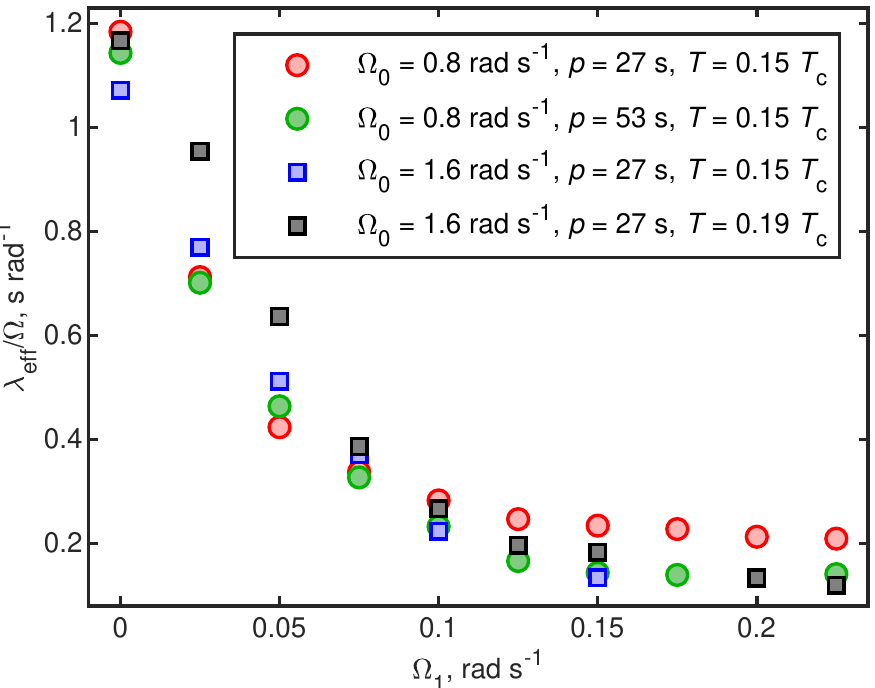}
\caption{\textbf{Effect of librating drive on the NMR frequency shift.} The steady state normalized ground level frequency shift $\lambda_{\rm eff}/\Omega$ as a function of the excitation amplitude $\Omega_1$ at $P=15.7$ bar, for different mean angular velocities $\Omega_0$, excitation periods $p$, and temperatures $T$. In all measured cases the normalized frequency shift saturates for excitation amplitudes $\Omega_1 \gtrsim 0.15$~rad~s$^{-1}$.}
\label{fig:freq-satur}
\end{figure}

The effect of vortices on the shape of the Bose-Einstein condensate trap and on the nuclear magnetic resonance (NMR) precession frequency appears via modification of the equilibrium texture. Vortex contribution to textural energy may be written as \cite{Thuneberg2001S,Eltsov2011S,Kopu2007S}
\begin{equation}
 F_{\mathrm{v}} = \frac{2}{5} a_{\mathrm{m}} H^{2} \frac{\lambda_{\mathrm{v}}}{\Omega} \int \mathrm{d}^3 r \frac{\left( \boldsymbol{\omega}_{\mathrm{v}} \cdot \hat{\mathbf{l}} \right)^2}{\omega_{\mathrm{v}}},
\end{equation}
where $a_{\mathrm{m}}$ is the magnetic anisotropy parameter, $H$ is the magnitude of the magnetic field, $\boldsymbol{\omega}_{\mathrm{v}} = \frac{1}{2} \langle \nabla \times \mathbf{v}_{\mathrm{s}} \rangle$ is the spatially averaged vorticity, $\hat{\mathbf{l}}$ is the orbital anisotropy vector, $\Omega$ is the angular velocity, and $\lambda_{\mathrm{v}} \propto \Omega$ is a dimensionless parameter characterizing vortex contribution to the textural energy. The vortex effect is a sum of two contributions. Namely, $\lambda_{\mathrm{v}} = \lambda_{\mathrm{f}} + \lambda_{\mathrm{c}}$, where $\lambda_{\mathrm{f}}$ originates from the orienting effect of the superflow around the vortex core, and $\lambda_{\mathrm{c}}$ is the vortex core contribution. When vortices are not completely polarized along the axis of rotation, the vorticity may be written as $\boldsymbol{\omega}_{\mathrm{v}} = \boldsymbol{\Omega} + \boldsymbol{\omega}_{\mathrm{v}}'$, where $\boldsymbol{\Omega}$ is the equilibrium part and $\boldsymbol{\omega}_{\mathrm{v}}'$ is a random part with $\langle \boldsymbol{\omega}_{\mathrm{v}}' \rangle = 0$. The equilibrium $\lambda_{\mathrm{v}}$ is then replaced by an effective value \cite{Eltsov2011S}
\begin{equation} \label{eq:lambda_eff}
 \lambda_{\mathrm{eff}} = \lambda_{\mathrm{v}} \frac{1 + ( \omega_{\mathrm{v\parallel}}/\Omega )^2 - ( \omega_{\mathrm{v\perp}}/\Omega )^2}{\sqrt{1 + ( \omega_{\mathrm{v\parallel}}/\Omega )^2 + 2( \omega_{\mathrm{v\perp}}/\Omega )^2}},
\end{equation}
where $\omega_{\mathrm{v\parallel}}$ and $\omega_{\mathrm{v\perp}}$ are the averaged vorticities of the random part along the axis of rotation and perpendicular to it, respectively. The experimental determination of $\lambda_{\mathrm{eff}}$ and its effect on the energy levels in the magneto-textural trapping potential are discussed in detail in Ref.~\citenum{Eltsov2011S}. Eq.~\eqref{eq:lambda_eff} also qualitatively agrees with the experimental observation, Fig.~\ref{fig:freq-satur}, that the observed $\lambda_{\rm eff}/\Omega$ decreases for increasing amplitude of excitation as the deviation of vortex lines from strictly vertical alignment is expected to grow.

\begin{figure} [tb]
\includegraphics[width=0.9\linewidth]{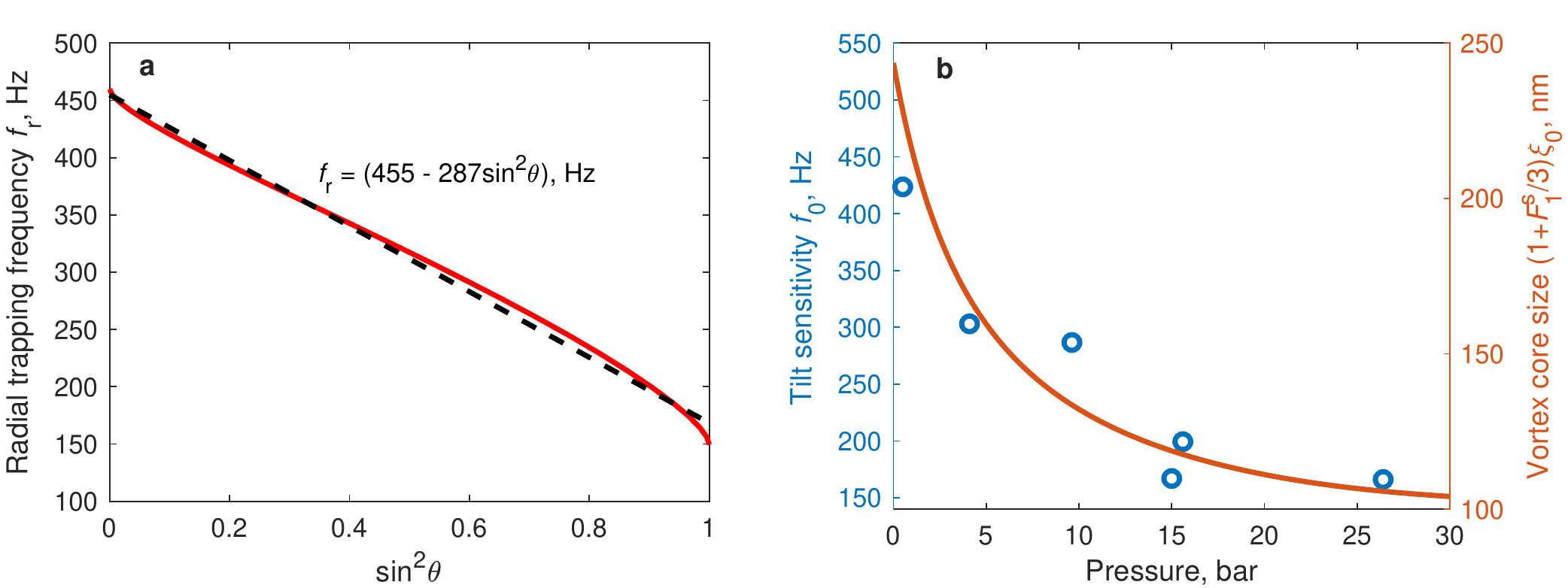}
\caption{\textbf{Frequency shift vs. vortex tilt angle.} {\bf a}, The calculated shift of the magnon condensate ground level at 832$\,$kHz Larmor frequency (red line) is approximately a linear function of $\sin^2 \theta$, where $\theta$ is the averaged vortex tilt angle. The dashed line is a linear fit to the numerically calculated frequency shift using experimentally determined value for $(\lambda_{\rm v}/\Omega)|_{\theta = 0}$ at $9.6\,$bar pressure. {\bf b}, As a function of pressure (here 832$\,$kHz NMR frequency), we find that the tilt sensitivity $f_0$ in Eq.~\eqref{eq:sin2theta} scales roughly with the vortex core size $(1+F_1^{\rm s}/3)\xi_0$, where $F_1^{\rm s}$ is the first symmetric Fermi-liquid parameter and $\xi_0$ is the zero-temperature coherence length.}
\label{fig:prefactor_analysis}
\end{figure}

The ground state frequency of the magnon BEC is determined by the order-parameter texture, which is affected by the vortex distribution and orientation. For a spiral vortex wave with the amplitude $a$ and wave vector $k_{\mathrm{z}}$ the tilt from the $z$ axis is
\begin{equation}
\theta = \arcsin \frac{a}{\sqrt{a^2 + k_{\mathrm{z}}^{-2}}} \ .
\label{thLexpr}
\end{equation}
Based on numerical calculation of the magnon BEC ground state in the model of uniform vortex tilt \cite{Kopu2007S}, the ground level frequency, see Fig.~\ref{fig:prefactor_analysis} is shifted by
\begin{equation} \label{eq:sin2theta}
\Delta f \approx -f_0 \sin^2\theta\,,
\end{equation}
where $f_0 \sim 100\,$Hz is a pressure-dependent parameter describing the tilt sensitivity, determined numerically. An example fit is shown in Fig.~\ref{fig:prefactor_analysis}. Although our model neglects the contribution of the global flow along the $z$ axis, naturally present for inertial waves, we expect this correction to remain small in the low-temperature limit where the superfluid density anisotropy, and therefore the texture-orienting effect of the counterflow ${\bf v}_{\rm n} - {\bf v}_{\rm s}$, where ${\bf v}_{\rm n}$ and ${\bf v}_{\rm s}$ are the superfluid and normal fluid velocities, respectively, vanishes \cite{PhysRevB.46.13983S}.


\begin{figure} [bt]
\includegraphics[width=0.7\linewidth]{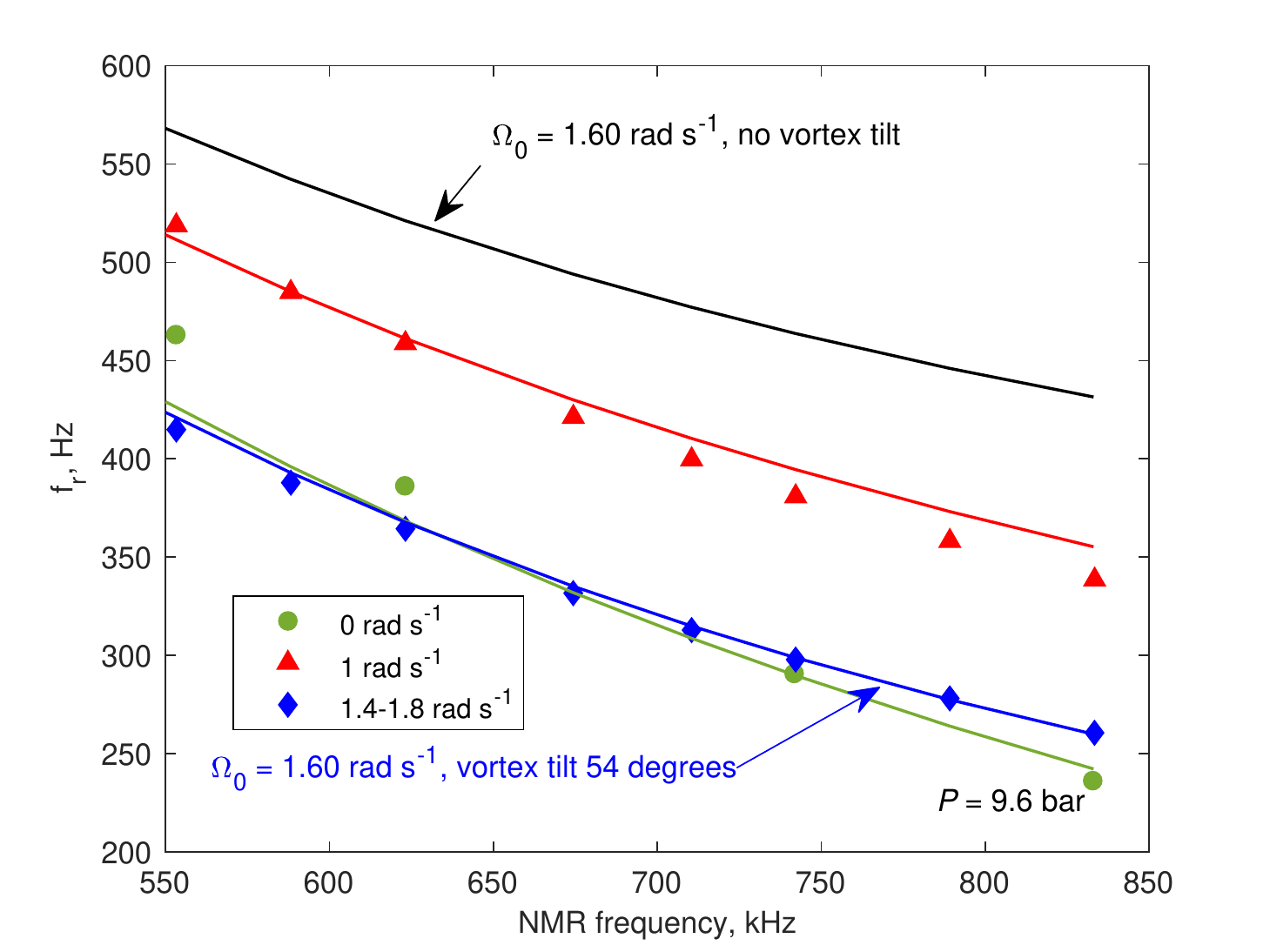}
\caption{\textbf{Textural trapping potential as a function of NMR frequency.} For steady angular velocity (straight vortices) the radial magnon BEC trapping frequency $f_{\rm r}$ (green and red symbols and lines) monotonously decreases for higher NMR frequency and increases for higher rotation velocity (higher vortex density). The experimental data during the librating drive (blue diamonds) is well fit with a model of uniformly tilted vortices (blue line), where the trap frequency is noticeably lower than without the tilt (black line).}
\label{fig:spectra-fit}
\end{figure}

Using the result of the fit shown in Fig.~\ref{fig:prefactor_analysis} on the frequency shift, seen for the upper spectrometer in Fig.~2{\bf a} in the main text, we estimate the average vortex tilt angle $\theta \approx 50^\circ \pm 10^{\circ}$. The uncertainty in the numerical calculations arises from the arbitrary orientation of the tilted vortex with respect to the radial direction. Our calculations also reproduce the experimentally observed dependence of $\Delta f$ on the NMR frequency $f$, Fig.~\ref{fig:spectra-fit}, and, as we will see later, agree with an independent estimate of the vortex tilt angle using a phenomenological model to describe the flow of energy towards the smallest scales.

\subsection*{\label{s:pump} Model of pumping}

In line with our experimental observations, we consider a model where we pump energy to a system of quantized vortices by dragging them from one end. Typically, if vortices were pinned in place at the boundary, they would be acted upon by mutual friction and by the Magnus force arising from motion of vortex lines with respect to the underlying normal and superfluid flow fields. However, as we will see, the total force in this case would far exceed the theoretical maximum under our experimental conditions. Instead, each vortex then feels a constant force equal to vortex tension, and the vortices form a quantum boundary layer with 'partial slip' boundary condition. We shall now consider this model in more detail.

The angular momentum of the superfluid rotating as a solid body with angular velocity $\Oms$ can be calculated as
\begin{equation}
 \mathcal{L} = 2 \pi L \rho_{\rm s} \int_0^R (\Oms r) r^2 {\rm d}r = \frac{1}{2} M R^2 \Oms\,,
\end{equation}
where $R$ is the radius of the cylindrical container, $L$ is the height of the container, $\rho_{\rm s}$ is the superfluid density, $M = \pi R^2 L \rho_{\rm s}$ is the total mass of the superfluid, and $\Oms$ is the (solid-body-like) angular velocity of the superfluid.

Assuming that vortex ends are pinned to the boundary and that $\Oms \neq \Omega$, where $\Omega$ is the angular velocity of the container, each vortex is acted upon by mutual friction $F_{\rm \alpha}$ and by the Magnus force $F_{\rm M}$. When $|{\bf F}_{\alpha} + {\bf F}_{\rm M}| > T_{\nu}$, where the vortex tension
\begin{equation}
 T_{\nu} = \rho_{\rm s} \kappa \nu_{\rm s}
\end{equation}
corresponds to tilting a vortex initially perpendicular to the surface parallel to it, the system forms a quantum boundary layer where the total force per vortex is limited in magnitude to $T_{\nu}$ \cite{PhysRevB.32.171S}. Here
\begin{equation}\label{OmegaC}
\nus = \frac{\kappa\Lambda}{4\pi}\,, \quad \Lambda \approx\ln
\frac{\ell}{a_0}\,,
\end{equation}
$\kappa \approx 6.6\cdot 10^{-4}\,$cm$^2$/s is the circulation quantum,  $\ell$ is the mean intervortex distance, and $a_0 \sim 0.1\,\mu$m is the vortex core radius. The critical angular velocity difference $\Delta \Omega = |\Omega - \Oms|$ at which point $|{\bf F}_{\alpha} + {\bf F}_{\rm M}|$ exceeds $T_{\nu}$ at a distance $r_0$ away from the center of rotation can be estimated as
\begin{equation}
 \Delta \Omega = \frac{T_{\nu}}{L \rho_{\rm s} \kappa r_0}\,.
\end{equation}
For $r_0 \sim 1$~mm we get $\Delta \Omega \sim 3 \cdot 10^{-4}$~rad$\,$s$^{-1}$, which is much smaller than the experimental changes in $\Omega$ and therefore we can safely use $T_{\nu}$ as an estimate of the force acting on the vortices at any instant.

The total torque acting on the superfluid, assuming each vortex is pulled with force $\beta T_{\nu}$, where we allow for a fitting parameter $\beta \sim 1$ to characterize vortex pulling efficiency, is
\begin{equation}
 \mathcal{T} = 2 \pi \beta \int_0^{R} T_{\nu} n_{\rm v} r^2 {\rm d}r = \beta \frac{4 \pi }{3} \frac{T_{\nu} R^3 \Oms}{\kappa} = \beta \frac{4 \pi }{3} \rho_{\rm s} \nu_{\rm s} R^3 \Oms\,.
\end{equation}
Here $n_{\rm v} = 2\Oms/\kappa$ is the density of vortices. Since the direction of the applied force depends on ${\rm sign} (\Omega - \Oms)$, we have $\dot{\mathcal{L}} = {\rm sign} (\Omega - \Oms) \mathcal{T}$. Solving for $\dot{\Omega}_{\rm s}$, we get
\begin{equation}
\frac{{\rm d} \Oms}{{\rm d} t} = \frac{2}{M R^2} \mathcal{T} = \beta {\rm sign} (\Omega - \Oms)  \frac{8 \nu_{\rm s}}{3 L R} \Oms \equiv \beta {\rm sign}(\Omega-\Oms) \tau_{\rm s}^{-1} \Oms \,,
\label{eq:Omegadot}
\end{equation} 
where $\taus = 3RL/(8\nus) \approx 4\cdot10^3\,{\rm s}$. For a step change in $\Omega: \Omega_0 \rightarrow \Omf$ at $t=0$ this equation has the solution
\begin{equation}
\Oms(t) = \Omega_0 \exp\left[\beta {\rm sign}(\Omf-\Omega_0) \frac{t}{\taus} \right] \,.
\end{equation}
We find that in this case $\Oms(t)$ reaches $\Omf$ in a finite time
\begin{equation}
\tf = \beta^{-1} \taus \left| \ln \frac{\Omf}{\Omega_0} \right| \,.
\label{tf}
\end{equation}
For the solution to remain physical, we impose that $\Oms(t>\tf) = \Omf$. We find that Eq.~\eqref{tf} is in good agreement with the measured $t_{\rm g}(\Omf) - t_{\rm g}(\Omega_0)$ (see Fig.~4 in the main text), which is a measure of the energy stored in the global flow. We note that for a small change $|\Omega_0 - \Omf|$, the time scale $\tf$ is much smaller than the time scale of the exponential change, i.e. $\tf \ll \taus$. Therefore, the change of the angular velocity of the system of vortices is roughly linear in time, i.e.
\begin{equation} \label{eq:OmsLinear}
 \Oms \sim \Omega_0 \left[ 1 + {\rm sign}(\Omf - \Omega_0) \frac{t}{\taus} \right]\,.
\end{equation}

\begin{figure*}[t]
\includegraphics[width=1\linewidth]{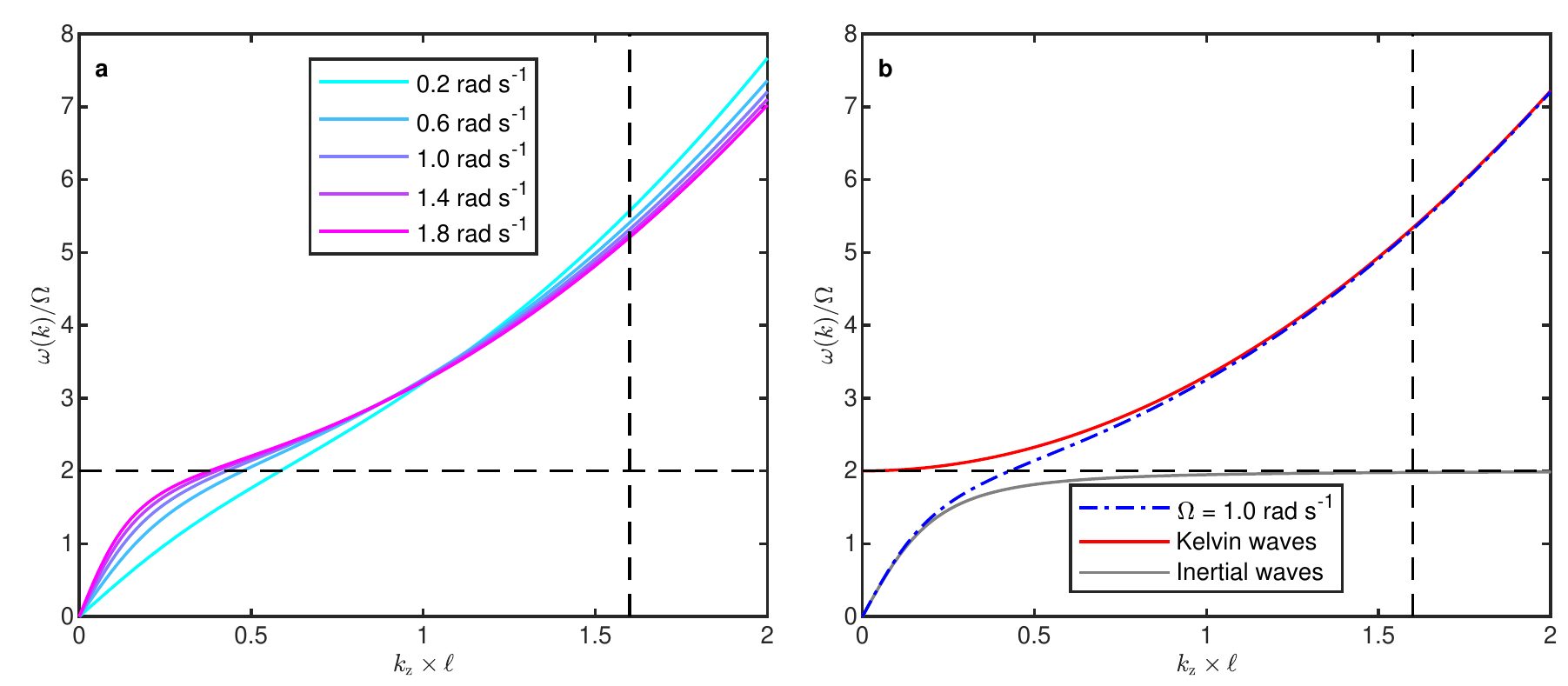}
\caption{\label{fig:vortspec} \textbf{Dispersion relation for vortex waves in a rotating cylinder.} \textbf{a} In the experimental range the dispersion relation of vortex waves has weak dependence of rotation velocity. \textbf{b} The full dispersion relation (blue dash-dotted line) interpolates smoothly between the Kelvin-wave (solid red line) and inertial wave (solid gray line) regimes. The crossover takes place at roughly $0.5\times k_{\mathrm{z}}\ell$ at typical angular velocity $1$~rad/s. For both plots the horizontal dashed black line shows the inertial wave cutoff frequency $2 \Omega$, and the vertical dashed black line shows the experimental estimation for the start of the KW cascade, see main text.}
\end{figure*}

In a coordinate system rotating with $\Omf$, the free energy per unit mass of the rotating superfluid is
\begin{equation}
 F = \frac{K - \Omf \mathcal{L}}{M} = \frac{\frac{1}{4}M R^2 \Oms^2 - \frac{1}{2} MR^2 \Oms \Omf}{M} = -\frac{R^2}{4} (\Oms^2 - 2 \Oms \Omf)  \,,
\end{equation}
where $K$ is the rotational kinetic energy of the superfluid in the inertial frame. Then, the power per unit mass pumped to the system by dragging vortices with $T_{\nu}$ becomes
\begin{equation} \label{eq:Wex}
 W_{\rm ex} = - \frac{{\rm d} F}{{\rm d}t} = - \frac{\partial F}{\partial \Oms} \frac{{\rm d} \Oms}{{\rm d}t} =\beta  \frac{4}{3} \frac{\nu_{\rm s} \Oms \left| \Omf - \Oms \right| R}{L} \,.
\end{equation}

We estimate the power pumped into the system during an oscillating drive (a triangle wave from $\Omega_0 - \Omega_1$ to $\Omega_0 + \Omega_1$ with period $p$)  by assuming that any additional energy terms arising from the angular acceleration of the rotating frame can be neglected (i.e. the rate of change for $\Omega$ is slow). Moreover, we assume that during one period of oscillation $\Oms$ remains approximately constant, $\Oms \approx \Omega_0$. With these assumptions, Eq.~\eqref{eq:Wex} remains a valid estimate for power pumped during the librating drive. Averaging Eq.~\eqref{eq:Wex} in time with these constraints gives the average power pumped into the superfluid during the librating drive
\begin{equation}
\left\langle W_{\rm ex}^{\rm osc} \right\rangle = \beta \frac23 \frac{\nus \Omega_0 \Omega_1 R}{L} \,.
\label{Wosc}
\end{equation}

\subsection*{Vortex waves in rotating superfluid}

The spectrum of waves for quantized vortices in a rotating cylinder at finite temperature $T$,
\begin{equation}\label{OmegaA} \omega(\vec k,T) \equiv  \omk + i \gak \,,
\end{equation}
was calculated in Ref~\cite{hendersonbarenghiS}. For the axially symmetric modes $\vec k = (k_{r},0,k_{z})$ the spectrum at $T=0$ where the dissipation vanishes and $\gak=0$ is given by
\begin{equation} \label{vortspecT0}
 \omega^2(\vec k, 0)=\omk^2 = \frac{(2\Omega+\nus k_{z}^2) [2\Omega+\nus(k_{z}^2+k_{r}^2)]}{1+(k_{r}/k_{z})^2}\ .
\end{equation}
Owing to the boundary condition $v_{r}(R) = 0$ the radial wave vector satisfies $J_1(k_{r} R) = 0$. The lowest radial mode has $k_{r} = k_{\mathrm{R}} = 3.8317/R$, thus fixing $k_{\mathrm{r}} = 13.1\,$cm$^{-1}$. As can be seen from Eq.~\eqref{vortspecT0}, in the classical $\nu_{\rm s} \rightarrow 0$ limit the spectrum transforms to the inertial-wave spectrum
\begin{subequations}\label{Omega1}\begin{equation}
\omk\IW = \frac{2\Omega}{\sqrt{1+(k_{r}/k_{z})^2}}\,.
\label{om-IW}
\end{equation}
With the typical excitation frequency $\omex = 0.23\,$s$^{-1}$ the IW spectrum gives $k_{\mathrm{z}}/k_{\mathrm{r}} \approx 0.07$, or $k_{\mathrm{z}} = 0.94$~cm$^{-1}$. For waves shorter than the inter-vortex distance (for which $k_{z} \gg k_{r}$) the spectrum takes the Kelvin-wave form
\begin{equation}
\omk\KW = 2\Omega + \nus k_{z}^2 \, .
\label{om-KW}
\end{equation}\end{subequations}
The transition between the two regimes takes place at $k_{z} \ell \sim 0.5$, Fig~\ref{fig:vortspec}.

At $T > 0$ the mutual friction leads to damping of the vortex waves, defined by  the imaginary part of frequency,  $\gak$. In  the Kelvin-wave limit it becomes
\begin{equation}
\gak\KW = \alpha(2\Omega + \nus k_{z}^2)= \alpha \omk\KW\,,
\label{gamma-KW}
\end{equation}
while in the inertial-wave limit
\begin{equation} \gak\IW = \frac{\alpha\Omega(k_{r}^2+2k_{z}^2)}{k_{r}^2+k_{z}^2}
= \alpha \left[\Omega + \frac{\left( \omk\IW \right) ^2}{4\Omega}\right]\, .
\label{gamma-IW}
\end{equation}

\begin{figure}
\includegraphics[width=0.5\linewidth]{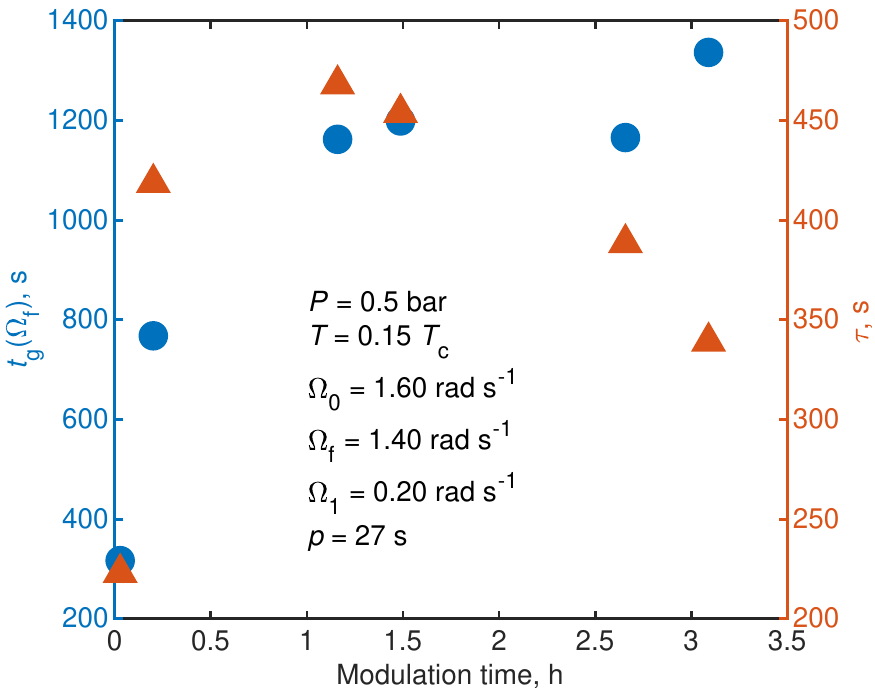}
  \caption{\textbf{Effect of modulation time on time scales.} The observed waiting times $t_{\rm g}(\Omf)$ (blue circles, left scale) remain constant after approximately one hour or modulations. The time constant $\tau$ (red triangles, right scale) of the exponential decay increases dramatically during the first few minutes, after which it saturates to a constant level and slightly decreases for the longest modulation times. Whether the non-monotonous behavior of $\tau$ is a real effect or originates from the random scatter remains unknown.}
  \label{fig:relax_vs_modtime}
\end{figure}

Here $\alpha$ is the dissipative mutual friction parameter, which in our experiments varies in the range $10^{-5} \div 10^{-3}$ \cite{PhysRevB.97.014527S}. For $T \lesssim 0.3 \, T_{\mathrm{c}}$ the reactive mutual friction parameter $\alpha' \approx \alpha^2 \ll 1 $ and can be neglected.

\subsection*{Cascade of inertial waves}

In the non-linear regime we can make crude estimates of the relevant time scales and energy cascade rates using dimensional reasoning in the spirit of Kolmogorov 1941 approach. In particular the lower limit for the time needed for developing of the turbulent cascade can be estimated as $t\sb{casc} \gtrsim 2L_{\rm out}/U$, where $L_{\rm out}$ is the outer scale of turbulence and $U$ is the velocity at this scale. We can estimate $L_{\rm out}$ as the diameter of the container, $L_{\rm out}\sim 2R$, which is also the wave length of the exited mode, since $k_{\mathrm{r}} \gg k_{\mathrm{z}}$ (formally $\lambda_{\rm ex} = 2\pi/k \approx 1.6 R$). Velocity $U$ is estimated from the root mean square velocity of the pinned vortex ends in the frame rotating with $\Omega_0$. The velocity of vortex ends is
\begin{subequations}
\begin{equation}
U\sb{est,1}\sim v_{\rm L,ex}^{\rm rms} = \sqrt{\langle [(\Omega(t) - \Omega_0) r]^2\rangle_{t,r}} = \frac{\Omega_1 R}{\sqrt{6}}\,,
\end{equation}
which gives $v_{\rm L,ex}^{\rm rms} \approx 0.025\,\mbox{cm\,s$^{-1}$}$. The timescale for the buildup of the cascade of vortex waves is thus
\begin{equation}
t\sb{casc,1}    \gtrsim   2L_{\rm out}/U\sb{est,1} \sim 50\,\mbox{s}\,,
\end{equation}
\end{subequations}
which is comparable to the observed time of the development of the signal, $t\sb{dev}\simeq 100\,$s. Buildup of the turbulent steady state over all length scales can take significantly longer, Fig.~\ref{fig:relax_vs_modtime}.

Alternatively, we can estimate the turbulence buildup time from the energy flux through the cascade. For weak IW turbulence, the energy (per unit mass) spectrum is anisotropic \cite{PhysRevE.68.015301S} if $k_{\mathrm{z}} < k_{\mathrm{r}}$
\begin{equation}
\mathrm{d} E_{\rm IW}(k_{\mathrm{r}},k_{\mathrm{z}}) \sim \sqrt{\varepsilon_{\rm IW} \Omega}\, k_{\mathrm{r}}^{-5/2} k_{\mathrm{z}}^{-1/2} \, \mathrm{d} k_{\mathrm{r}}\, \mathrm{d} k_{\mathrm{z}}\ ,
\label{dEIW1}
\end{equation}
where $\varepsilon_{\rm IW}$ is the energy flux per unit mass. The total energy is determined by the smallest $k_{\mathrm{r}} = k_{\mathrm{R}}$ and the largest $k_{\mathrm{z}} \sim k_{\mathrm{R}}$:
\begin{equation}
E_{\rm IW} \sim  \sqrt{\varepsilon_{\rm IW} \Omega}\, k_{\mathrm{R}}^{-1} \,.
\label{EIW}
\end{equation}
On the other hand, for $k_{\mathrm{z}} > k_{\mathrm{r}}$ the energy spectrum becomes nearly isotropic and we can write
\begin{equation}
\mathrm{d} E_{\rm IW}(k) \sim \sqrt{\varepsilon_{\rm IW}\Omega}\, k^{-2} \, \mathrm{d} k \, .
\label{dEIW2}
\end{equation}
Here the energy is determined by the smallest $k \sim k_{\mathrm{R}}$ and we again arrive to Eq.~\eqref{EIW}.

For superfluids at low temperatures, the dissipation takes place at length scales smaller than the inter-vortex distance. Thus, in the IW regime the energy stored in IWs then follows the differential equation
\begin{equation}
\frac{\mathrm{d} E_{\rm IW}}{\mathrm{d} t} = W_{\rm ex} - \varepsilon_{\rm IW} \,.
\label{EIW-dot}
\end{equation}
For weak turbulence we have from Eq.~\eqref{EIW}
\begin{equation}
\varepsilon_{\rm IW} =  \frac{(k_{\mathrm{R}} E_{\rm IW})^2}{\Omega} \,.
\label{epsIW}
\end{equation}
However, from the observed vortex tilt angle $\sim 50^{\circ}$ we may expect that non-linearities are strong. In this case the energy flux approaches strong hydrodynamics turbulence limit $ \varepsilon_{\rm HD} \sim k_R E_{\rm HD}^{3/2} \,.$ We suggest a simple  interpolation formula between these two regimes
\begin{equation}
\varepsilon_{\rm IW} = \,\frac{(k_{\mathrm{R}} E_{\rm IW})^2}{\Omega + k_{\mathrm{R}} \sqrt{E_{\rm IW}}}.
\label{epsIW2}
\end{equation}

The steady state solution of Eq.~\eqref{EIW-dot} is $\varepsilon_{\rm IW} = W_{\rm ex}$. For typical parameters of librating motion, $\Omega_0 = 1.6\,$rad~s$^{-1}$ and $\Omega_1 = 0.2\,$rad~s$^{-1}$, we get $W_{\rm ex} \approx 2\cdot 10^{-6}\,$cm$^2$~s$^{-3}$ from Eq.~\eqref{Wosc}. Equating it to the energy flux in Eq~\eqref{epsIW2}, we obtain $E_{\rm IW} \approx 10^{-4}\,$cm$^2$~s$^{-2}$. Thus, in Eq.~\eqref{epsIW2} the term $k_{\mathrm{R}} \sqrt{E_{\rm IW}} \sim 0.1\,{\rm rad\,s^{-1}} \ll \Omega_0$ and the energy cascade corresponds to that in Eq.~\eqref{epsIW}, which corresponds to the weak turbulence regime. In this case we get another estimate for the characteristic velocity by calculating the steady state value of $E_{\rm IW}$ by plugging Eq.~\eqref{epsIW} into Eq.~\eqref{EIW-dot} and calculating
\begin{equation}\label{estU-2}
 U\sb{est,2} \approx  \sqrt{2\,E_{\rm IW}} \approx 1.4 \cdot 10^{-2}\,\mbox{cm~s$^{-1}$}\,.
 \end{equation}
We find that $U \ll \Omega_1 R$ and thus vortex lines in bulk cannot follow the motion of the container walls. The estimate \eqref{estU-2} results in the cascade buildup time
\begin{equation}\label{casc-2}
  t\sb{casc,2} \gtrsim 2 L_{\rm out}/U\sb{est,2} \sim 80\,\mathrm{s}\,,
\end{equation}
in good agreement with the experimentally observed developing time $t\sb{dev}\simeq 100\,$s.

Let us now estimate the average tilt angle caused by inertial waves. The tilt angle at the scale $k$ is $\sin \theta \sim (U/\omega_{k}) k$, where $U/\omega_{k}$ corresponds to the wave amplitude $a$ in Eq.~\eqref{thLexpr}. For vortex waves with random phases one has to sum $\sin^2 \theta \sim (E/\omega_{k}^2) k^2$. For $k > k_{\mathrm{R}}$ we have $\omega_{k} \sim \Omega$ and from Eq.~\eqref{dEIW2} we get
\begin{equation}
    \mathrm{d} \sin^2 \theta \sim \frac{\mathrm{d} E}{\Omega^2} k^2 \sim \frac{\sqrt{\varepsilon_{\rm IW} \Omega}}{\Omega^2}\, \mathrm{d} k \,.
\end{equation}
The integral is determined by the largest $k$ and we get
\begin{equation}
\sin^2 \theta \sim \varepsilon_{\rm IW}^{1/2} \Omega^{-3/2} k_\ell \sim
E_{\rm IW} \Omega^{-2} k_{\mathrm{R}} k_\ell \,.
\label{thIW}
\end{equation}
Here $k_\ell = 2 \pi \ell^{-1}$ is the wave vector corresponding to the inter-vortex length scale. For $\varepsilon_{\rm IW} = W_{\rm ex} = 2\cdot 10^{-6}\,$cm$^2$~s$^{-3}$ estimated above we obtain $\sin^2\theta_{\rm IW} \approx 0.06$. This contribution alone disagrees with NMR measurements, but we will see that KWs provide a larger contribution. We note that while the estimated buildup time for IW energy cascade agrees with our experimental observations, the observed change in the vortex polarization well exceeds our prediction. Likely, we either underestimate the effect of IWs on vortex polarization or generate KWs by dragging vortices along the surface \cite{PhysRevB.97.014527S}. We further note that the observed saturation of vortex polarization does not fully determine the state of the system since the dynamics resulting from stopping the drive saturate only after about an hour, Fig.~\ref{fig:relax_vs_modtime}. We take this as the buildup time for KW cascade, in agreement with our phenomenological model as we sill see later.

The part of the spectrum with $k_{z} < k_{r} \sim k_{\mathrm{R}}$ has only a minor contribution to $\sin^2 \theta$. Indeed, from Eq.~\eqref{dEIW1} we find
\begin{widetext}
 \begin{equation}
  \mathrm{d} \sin^ 2 \theta \sim \frac{\mathrm{d} E}{\omega_k^2} k_{z}^2 \sim \sqrt{\varepsilon_{\rm IW} \Omega} \,\frac{1 + (k_{r}/k_{z})^2}{(2\Omega)^2}\, 
k_{r}^{-5/2} k_{z}^{-1/2} k_{z}^2 \, \mathrm{d} k_{r} \, \mathrm{d} k_{z}
\sim \varepsilon^{1/2} \Omega^{-3/2} k_{r}^{-1/2} k_{z}^{-1/2}\, \mathrm{d}k_{r} \, \mathrm{d}k_{z}\,.
 \end{equation}
\end{widetext}
The integral is determined by large $k_{z} \sim k_{r} \sim k_{\mathrm{R}}$ and thus
\begin{equation}
\sin^2 \theta \sim \varepsilon_{\rm IW}^{1/2} \Omega^{-3/2} k_{\mathrm{R}},    
\end{equation}
which is significantly smaller than the contribution in Eq.~\eqref{thIW}.

We note that the {\it critical height for two-dimensionalization of the flow} for typical rotation velocity $\Omega = 1.6\,$rad$\,$s$^{-1}$ in the experiments, from Ref.~\cite{ALEXAKIS20181S},
\begin{equation}
 L_{\rm c} \sim \Omega \varepsilon_{\rm IW}^{-1/3} k_{\rm in}^{-5/3} \approx 1.7\,{\rm cm}\,,
\end{equation}
where $k_{\rm in} = \sqrt{k_{r}^2 + k_{z}^2} \approx 13.1\,$cm$^{-1}$ corresponds to the wave vector for the first axially symmetric inertial wave mode (see main text for details), is much smaller than the container height $L$. In this case there is coupling between the slow and fast manifolds and the expected flow is effectively three-dimensional. The coupling between the manifolds leads to strictly forward cascade (i.e. towards smaller scales) and should eventually transfer energy to the KW regime.

\subsection*{Cascade of Kelvin waves}

Up to our knowledge, the Kelvin-wave cascade in rotating vortex array has not been considered theoretically. We will thus use results for the single-vortex case in the absense of rotation \cite{PhysRevB.91.144501S,KW_AmplitudeS} as a basis of our analysis. Here the leading KW interaction is taken from the effective 4-wave theory of the $1 \Rightarrow3$ type. In the single vortex case the gap in the KW spectrum, Eq.~\eqref{om-KW}, drops out from the dispersion relation. Thus, we expect that our analysis will provide us with qualitative (and possibly semi-quantitative) description of the rotating case. The L'vov-Nazarenko KW energy spectrum (per unit vortex length and per superfluid density) is \cite{Lvov2010S}
\begin{equation}
\mathcal{E}_{k} = c\Sb{LN} \Lambda \kappa \epsilon_{\rm KW}^{1/3} \Psi^{-2/3} k^{-5/3} \,,
\label{dEKW}
\end{equation}
where $c\Sb{LN} \approx 0.304$ is the L'vov-Nazarenko factor and $\epsilon_{\rm KW}$ is the the energy flux per unit vortex length and per unit superfluid density. The dimensionless factor 
\begin{equation}
\Psi = \frac{8\pi}{\Lambda \kappa^2} \int_{k_{\rm start}}^{\infty} \mathcal{E}_k {\rm d}k 
\end{equation}
can be thought as the mean square of the vortex tilt angle, while the KW cascade starts at wave vector $k_{\rm start} \sim \ell^{-1}$. Solving for $\Psi$ gives
\begin{equation} \label{eq:Psi}
 \Psi = \frac{(12 \pi c\Sb{LN})^{3/5}\epsilon_{\rm KW}^{1/5}}{\kappa^{3/5}k_{\rm start}^{2/5}}\,.
\end{equation}
The energy contained in the system of KWs for $N_{\rm v} = \pi R^2 \ell^{-2}$ vortices in a cylindrical container with radius $R$ and length $L$ can be written as
\begin{widetext}
\begin{equation}
\mathcal{E}_{\rm KW} = \rho_{\rm s} L_{\rm tot} \int_{k_{\rm start}}^{\infty} \mathcal{E}_{k} {\rm d}k = \rho_{\rm s} L_{\rm tot} \frac{3 c_{\rm LN}^{3/5} \Lambda}{2(12 \pi)^{2/5}} \kappa^{7/5} \epsilon_{\rm KW}^{1/5} k_{\rm start}^{-2/5} \,,
\label{EKW}
\end{equation}
\end{widetext}
where $L_{\rm tot} = N_{\rm v} L$ is the combined length of straight vortices. This allows rewriting Eq.~\eqref{dEKW} as
\begin{equation}
\mathcal{E}_k = \frac23 \frac{\mathcal{E}_{\rm KW}}{\rho_{\rm s} L_{\rm tot}} k^{-5/3} k_{\rm start}^{2/5}\,.
\label{dEKW1}
\end{equation}
Let us also write the energy cascade rate as a function of the total energy in the KW cascade $\mathcal{E}_{\rm KW}$
\begin{equation} \label{eq:epsilonKW}
 \epsilon_{\rm KW} = \frac{512 \, \pi^2}{27 \, c_{\rm LN}^3 \Lambda^5} \frac{k_{\rm start}^2}{\kappa^7} \left( \frac{\mathcal{E}_{\rm KW}}{\rho_{\rm s} L_{\rm tot}} \right)^5\,.
\end{equation}

\begin{figure}
\includegraphics[width=0.4\linewidth]{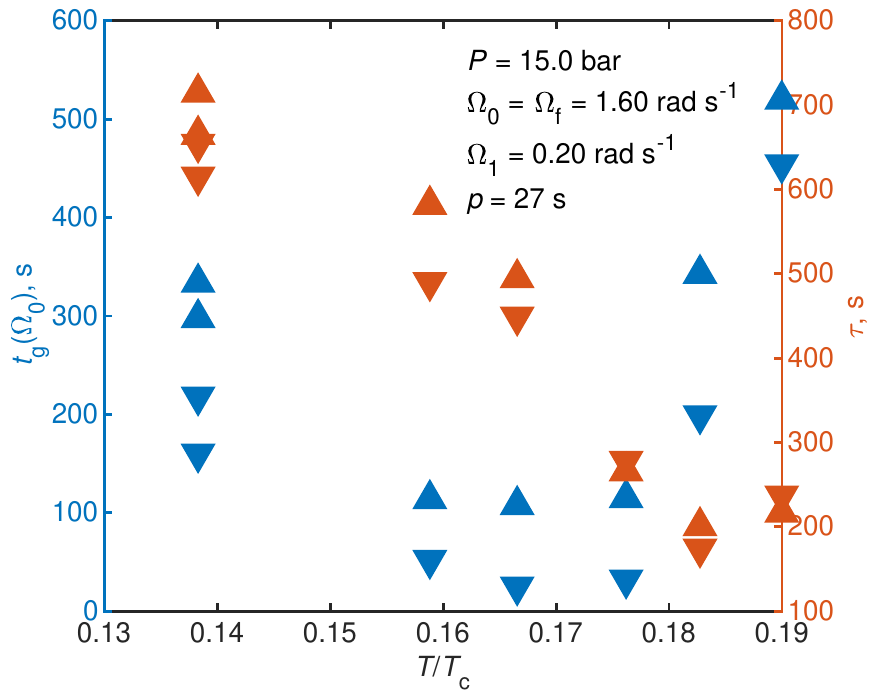}
\caption{\textbf{Effect of temperature on time scales.} In the plot the upwards-pointing triangles correspond to data measured with the upper NMR spectrometer, while the downwards-pointing triangles correspond to data measured with the lower NMR spectrometer. The observed waiting time $t_{\rm g}(\Omega_0)$ has a non-monotonous dependence on the temperature - it decreases sharply with decreasing temperature until $T \lesssim 0.18 T_{\mathrm{c}}$, saturates between $0.15 T_{\mathrm{c}} \lesssim T \lesssim 0.18 T_{\mathrm{c}}$, and increases again below $\sim 0.15 T_{\mathrm{c}}$. On the other hand the time constants $\tau$ for the exponential decay are observed to monotonically decrease with increasing temperature - as expected for increasing dissipation.}
\label{fig:t00_vs_temp}
\end{figure}

According to Ref.~\cite{Bou__2012S} the kinetic equation for the Kelvon occupation number $n_{k}$ in the presence of finite mutual friction dissipation can be written as
\begin{equation}
 \frac{\partial n_{{\bf k}}}{\partial t} = {\rm  St}_{\bf k} \{ n_{{\bf k}'} \} - \alpha \omega_{\rm KW}'(k) n_{\bf k}\,,
\end{equation}
where ${\rm St}$ is a collision term for the effective 4-wave interaction defined by Eq.~(5) in Ref.~\cite{Bou__2012S} and $\omega_{\rm KW}'(k) = \nu_{\rm s} k^2$ is the gapless KW dispersion relation. Since the energy (per unit length and per unit superfluid density) at wave vector $k$ is $\mathcal{E}_{k} = 2 \omega_{\rm KW}'(k) n_k$ (where the factor 2 comes from the assumption $n_{\bf k} = n_{\bf -k}$ which is not necessarily true for the rotating case), it follows that
\begin{equation} \label{eq:Ek}
 \frac{\partial \mathcal{E}_k}{\partial t} = 2 \omega_{\rm KW}'(k) {\rm St}_k - 2 \alpha \omega_{\rm KW}'(k) \mathcal{E}_k\,.
\end{equation}
The evolution of the total energy stored in the system of KWs can be obtained by integrating Eq.~\eqref{eq:Ek} over $k$
\begin{equation}
 \frac{\partial \mathcal{E}_{\rm KW}}{\partial t} = \rho_{\rm s} L_{\rm tot} \int_{0}^{\infty} \left[ 2 \omega_{\rm KW}'(k) {\rm St}_k - 2 \alpha \omega_{\rm KW}'(k) \mathcal{E}_k \right] {\rm d}k \equiv  M_{\rm tot} \varepsilon_{\rm IW} - 2 \alpha \rho_{\rm s} L_{\rm tot} \int_{k_{\rm start}}^{k_{\rm end}} \omega_{\rm KW}'(k) \mathcal{E}_k {\rm d}k\,,
\end{equation}
where $M_{\rm tot} =\rho_{\rm s} \pi R^2 L$ is the total mass of the superfluid. The cascade of KWs extends between $k_{\rm end}$ and $k_{\rm start} \ll k_{\rm end}$, and the input flux is equated to the energy flux from the inertial wave cascade. Substituting $E_{\rm KW} = \mathcal{E}_{\rm KW} M_{\rm tot}^{-1}$ on the left hand side and taking $\mathcal{E}_k$ from Eq.~\eqref{dEKW1}, we get
\begin{equation}
 \frac{\partial E_{\rm KW}}{\partial t} = \varepsilon_{\rm IW} - 2 \alpha \nu_{\rm s} k_{\rm start}^{2/3} k_{\rm end}^{4/3} E_{\rm KW} \equiv \varepsilon_{\rm IW} - 2 \alpha \omega_{\rm KW}'\left( k_\alpha \right) E_{\rm KW} \,.
 \label{EKW-dot}
\end{equation}
Comparing this result with the damping of a single KW mode, Eq.~\eqref{gamma-KW}, we find that the decay of the total energy in the KW cascade proceeds as if it took place at an effective length scale $k_\alpha = k_{\rm start}^{1/3} k_{\rm end}^{2/3}$. We note that this expression does not explicitly contain a term involving the energy flux $\epsilon_{\rm KW}$ since in this model the only energy sink is provided by mutual friction, which acts across all relevant length scales and is therefore not limited by the energy cascade rate. In the steady-state pumping we have $\varepsilon_{\rm IW} = W_{\rm ex}$, which for $W_{\rm ex} = 2\cdot 10^{-6}\,$cm$^2$~s$^{-3}$ and $\Omega=1.6\,$rad~s$^{-1}$ results in $E_{\rm KW} \sim 10^{-3}$cm$^2$~s$^{-2}$. To estimate the average tilt angle from KWs, we use the relation \cite{KW_AmplitudeS}
\begin{equation} \label{eq:tantheta}
 \left\langle \tan^2\theta \right\rangle \simeq \Psi\,.
\end{equation}
Using $\epsilon_{\rm KW} \sim \varepsilon_{\rm IW} = 2 \cdot 10^{-6}$~cm$^2\,$s$^{-3}$ and $k_{\rm start} \sim 1.6 \, \ell^{-1} \sim 70$~cm$^{-1}$ in Eq.~\eqref{eq:Psi}, we get an estimate $\theta \sim 57^\circ$, which is comparable with the angle extracted from the observed frequency shift during the steady state, Fig.~2{\bf b} in the main text.

Let us now turn our attention to the time scale of the relaxation. When $\varepsilon_{\rm IW} \rightarrow 0$ in Eq.~\eqref{EKW-dot}, the decay of energy becomes exponential with a rate
\begin{equation}
 \tau^{-1} = 2 \alpha \nu_{\rm s} k_\alpha^2 = 2 \alpha \nu_{\rm s} k_{\rm start}^{2/3} k_{\rm end}^{4/3}\,.
 \label{eq:tau}
\end{equation}
Since the crossover between the IW and the KW regimes takes place at $k \ell \sim 1$, Fig.~\ref{fig:vortspec}, we can assume that the length scale for the start of the cascade is fixed by the inter-vortex distance, $k_{\rm start} = C_{\rm start}\ell^{-1}$, where $C_{\rm start} \sim 1$. According to Ref.~\cite{Bou__2012S}, we can estimate the wave vector $k_{\rm end}$ at which the cascade is terminated as
\begin{equation}
 k_{\rm end} \approx \left( \frac{32 \pi}{3} \right)^{3/4} \frac{\sqrt{\nu'} \Psi}{\sqrt{\kappa} \left( \Lambda \sqrt{\alpha c_{\rm LN}} \right)^{3/2}} \ell^{-1} \,,
\end{equation}
where $\nu' \lesssim \kappa$ is the {\it effective kinematic viscosity} characterizing the energy dissipation rate in quantum turbulence. Previous experiments using $^3$He-B \cite{PhysRevB.85.224526S, PhysRevLett.96.035301S}, superfluid $^4$He \cite{PhysRevLett.100.245301S, PhysRevLett.99.265302S, PhysRevB.94.094502S}, as well as simulations \cite{PhysRevB.62.11751S, PhysRevA.67.015601S}, show that, depending on how the turbulence is generated, $\nu'$ may obtain quite different values, especially at low temperatures. Previously reported values range from $\sim 10^{-4}\kappa$ to $\sim 10^{-1} \kappa$. In general, smaller values for $\nu'$ are obtained for turbulence resulting from spin-down \cite{PhysRevB.85.224526S, PhysRevLett.100.245301S, PhysRevLett.99.265302S}, where vortices are (at least partially) polarized, and in simulations in the absence of reconnections (i.e. for wave turbulence) \cite{PhysRevA.67.015601S}. Theoretically, suppression of $\nu'$ has been suggested for polarized vortex tangles \cite{PhysRevB.76.024520S}. On the other hand, higher values of $\nu'$ are obtained in counterflow turbulence measurements, where the turbulence is homogeneous and isotropic \cite{PhysRevB.94.094502S, Skrbeke2018406118S}. 

\begin{figure} [tb]
\includegraphics[width=\linewidth]{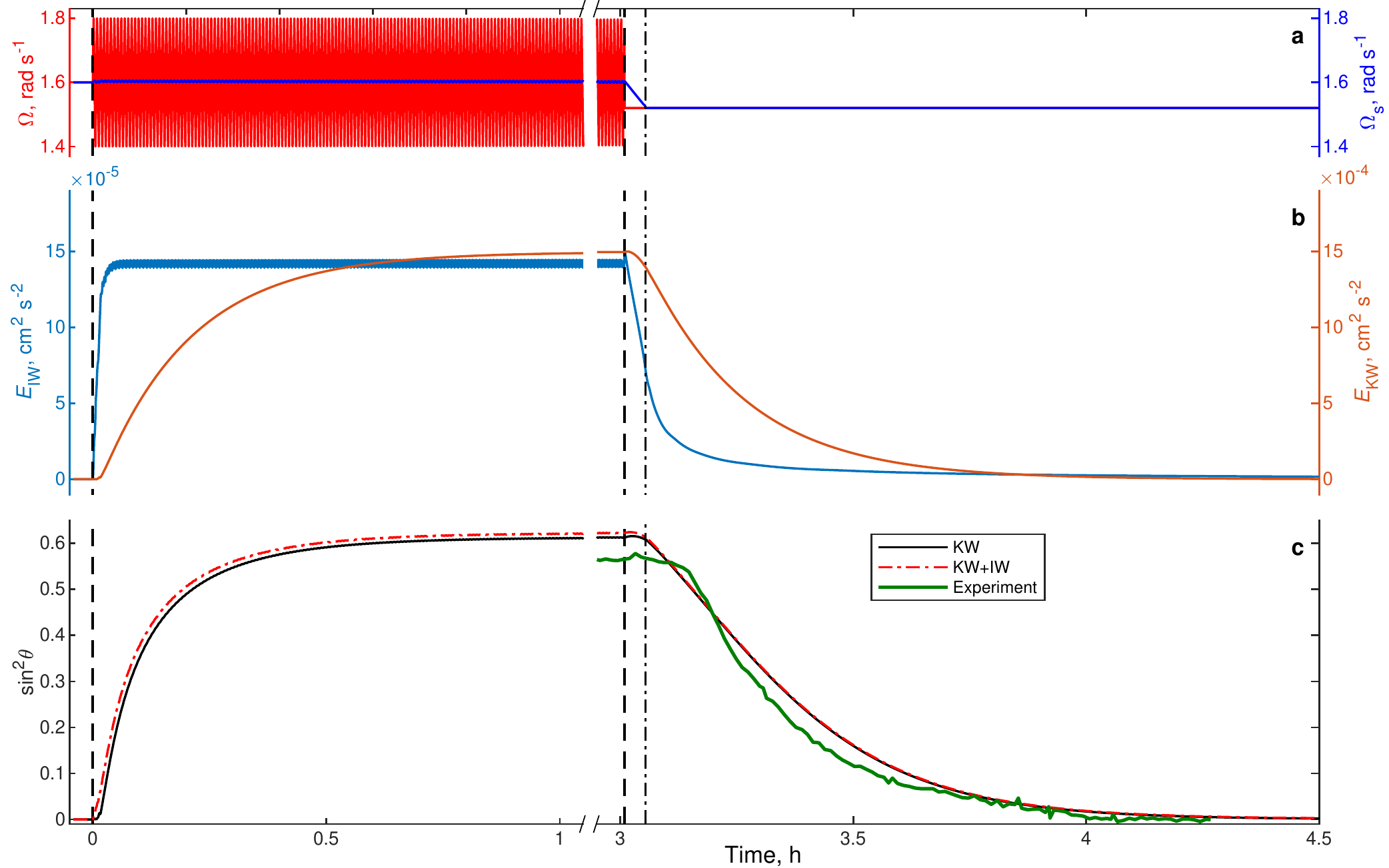}
\caption{\textbf{Numerical simulations - dynamics.} We numerically simulate the evolution of the system of Equations~\eqref{eq:system} using experimentally relevant drive $\Omega$, which is a trangle wave between 1.4$\,$rad$\,$s$^{-1}$ and 1.8$\,$rad$\,$s$^{-1}$ with period $p = 26 \frac{2}{3}\,$s. In the simulations the drive continues for three hours, after which it is stopped at 1.52$\,$rad$\,$s$^{-1}$ in line with the experiment presented in Fig.~2 in the main text. The values of $\mathcal{A}$ and $\beta$ used in simulations are within a few percent from the experimentally determined values for that measurement ($P=9.6\,$bar and $T = 0.13\,T_{\rm c}$). The initial conditions are $\Oms = 1.6\,$rad$\,$s$^{-1}$, and $E_{\rm IW} = E_{\rm KW} = 0$. The dashed vertical lines correspond to the beginning and to the end of the drive, and the vertical dash-dotted line shows the moment when $\Oms$ reaches $\Omf$. {\bf a} The superfluid angular velocity $\Oms$ (blue) remains almost constant in the beginning and during the drive (red). At the end of the modulation drive $\Oms$ changes to the final value in a nearly linear fashion, as expected from Eq.~\eqref{eq:OmsLinear}. The evolution of the energies $E_{\rm IW}$ and $E_{\rm KW}$ are shown in panel {\bf b}. When the modulation is turned on, $E_{\rm IW}$ develops to its final value within $\sim 100\,$s in agreement with the observed experimental time scale. For $E_{\rm KW}$ this takes significantly longer, about an hour, consistent with the experimental observation of the buildup time for $t_{\rm g}(\Omf)$, Fig.~\ref{fig:relax_vs_modtime}. The energies are converted to the average tilt angle in panel {\bf c}. The buildup time of $\sin ^2 \theta$ in simulations in the beginning of the modulations is longer than the experimental one. During the steady state, the corresponding vortex tilt angle is $\theta \approx 51^\circ$, in line with our estimates from the NMR frequency shift. After the modulation is stopped, $\sin ^2 \theta$ remains close to its steady state value approximately until $\Oms$ reaches $\Omf$. When $E_{\rm IW}$ has significantly decreased, exponential decay of $\sin^2\theta$ starts. The black line shows the contribution of KWs to the average tilt angle,  the red dash-dotted line shows the total angle calculated with Eq.~\eqref{eq:totalAngle}, and the solid green line is the experimental data from Fig.~2{\bf b} in the main text, converted to $\sin^2 \theta$ using Eq.~\eqref{eq:sin2theta} (see also Fig.~\ref{fig:prefactor_analysis}) and shifted in time such that the drive $\Omega$ reaches $\Omf$ at the same time in the experiment and simulations.}
\label{fig:numeric_sim}
\end{figure}

Plugging in $k_{\rm start}$ and $k_{\rm end}$ to Eq.~\eqref{eq:tau}, we get
\begin{equation}
 \tau^{-1} = \frac{16}{3} \frac{C_{\rm start}^{2/3} \kappa^{1/3} \nu'^{2/3} \Psi^{4/3}}{c_{\rm LN} \Lambda \ell^{2}} \,,
 \label{eq:tau2}
\end{equation}
which is, assuming constant $\nu'$, {\it independent} of temperature. Moreover, Eq.~\eqref{eq:tau2} depends linearly on the vortex density, $n_{\rm v} \propto  \ell^{-2} \propto \Omf$. Therefore, we can write
\begin{equation}
 \frac{1}{\tau \Omf}  = \frac{32}{3} \frac{C_{\rm start}^{2/3} (\nu'/\kappa)^{2/3} \Psi^{4/3}}{c_{\rm LN} \Lambda} \,.
 \label{eq:tau_constant}
\end{equation}
Setting $\Psi \simeq 1$ and $C_{\rm start} = 1.6$ extracted from experimental data and using the effective kinematic viscosity $\nu'$ as a free parameter, we estimate that at the lowest temperatures $\nu' \sim 10^{-6} \kappa$, Fig.~4 in main text. This value is two orders of magnitude below the lowest value reported previously in literature \cite{PhysRevB.85.224526S} and five orders of magnitude below values reported for homogeneous and isotropic turbulence \cite{PhysRevB.94.094502S}, likely due to high vortex polarization and low reconnection rate.

\subsection*{Total tilt angle}

We estimate the total tilt angle from both IWs and KWs by a simple interpolation formula
\begin{equation} \label{eq:totalAngle}
 \sin^2 \theta_{\rm tot} \sim \frac{1}{1 + \left( \Psi + 1.6 E_{\rm IW} k_{\rm R} \ell^{-1} \Oms^{-2} \right)^{-1}}\,,
\end{equation}
where we use $k_\ell \sim 1.6 \ell^{-1}$ (note that here $k_\ell$ is the same as $k_{\rm start}$) in Eq.~\eqref{thIW} as the transition length scale between the IWs and the KWs (see also main text). Using $\Psi \approx \tan^2 ( 57^{\circ})$ and $E_{\rm IW} \approx 10^{-4}\,$cm$^2$s$^{-2}$ from previous sections we get $\theta_{\rm tot} \sim 57^{\circ}$, which is essentially unaffected by the contribution from IWs.

\begin{figure} [tb]
\includegraphics[width=\linewidth]{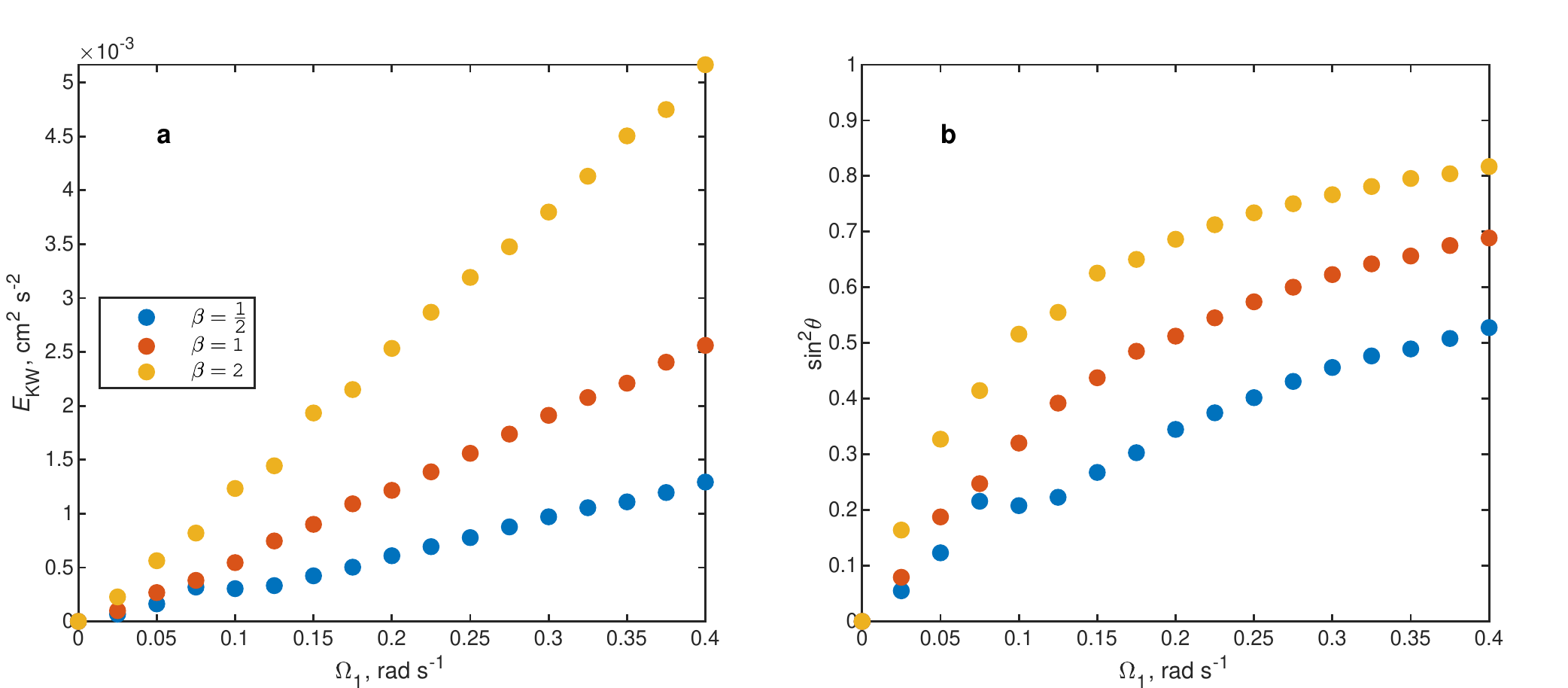}
\caption{\textbf{Numerical simulations - steady state.} We simulate the steady state behavior of the averaged tilt angle as a function of the modulation amplitude $\Omega_1$ by driving the vortices out of equilibrium with a trangle wave between 1.4$\,$rad$\,$s$^{-1}$ and 1.8$\,$rad$\,$s$^{-1}$ with period $p = 26 \frac{2}{3}\,$s for three hours. From the simulations, we extract $E_{\rm KW}$ (panel {\bf a}), which is then converted to $\sin^{2}\theta$ (panel {\bf b}). The simulations qualitatively reproduce the apparent saturation seen in the experiments, Fig.~\ref{fig:freq-satur}.}
\label{fig:numeric_sim_saturation}
\end{figure}

\subsection*{Numerical simulations of phenomenological model for energy stored in vortex waves}

In the simulations, we numerically solve a system of differential equations for $\Oms$, $E_{\rm IW}$, and $E_{\rm KW}$, resulting from Eqs.~\eqref{eq:Omegadot}, \eqref{eq:Wex}, \eqref{epsIW}, and \eqref{EKW-dot}:
\begin{equation}
\begin{dcases} \label{eq:system}
 \dot{\Omega}_{\rm s} &= 
 \begin{cases}
    \beta \tau_{\rm s}^{-1} {\rm sign}(\Omega - \Oms) \Oms & \text{if } t<\tf\,,\\
    0 & \text{otherwise,}
 \end{cases} \\
 \dot{E}_{\rm IW} &= \beta \frac{4}{3} \frac{{\nu_{\rm s} \Oms |\Omega - \Oms|} R}{L} - \frac{(k_{\rm R} E_{\rm IW})^2}{\Oms} \text{,} \\
 \dot{E}_{\mathrm{KW}} &= \frac{(k_{\rm R} E_{\rm IW})^2}{\Oms} - 2 \alpha \nu_{\rm s} k_{\alpha}^2 E_{\rm KW} = \frac{(k_{\rm R} E_{\rm IW})^2}{\Oms} - \mathcal{A} \Oms E_{\rm KW}\,\text{,}
\end{dcases}
\end{equation}
where dot refers to time derivative and $k_{\alpha}^{2} = \frac{1}{2}\mathcal{A}\Oms \alpha^{-1} \nu_{\rm s}^{-1}$ is the effective dissipation length scale for mutual friction extracted from experiments, Fig.~4 in the main text. We note that with this definition of $k_\alpha$, it follows that $\dot{E}_{\mathrm{KW}}$ does not depend on $\alpha$ directly but rather the timescale of the exponential decay is set by experimentally determined slope for $\tau^{-1} = \mathcal{A} \Omf$. Moreover, $k_{\alpha}$ remains $\sim$constant with time, while generally in decaying quantum turbulence characteristic length scales may depends on time \cite{Skrbeke2018406118S}. In our case the time-independent value of the inter-vortex distance maintained by rotation may in principle stabilize also other relevant length scales.

The response of the system of Eqs.~\eqref{eq:system} to a librating drive with experimental parameters is presented in Fig.~\ref{fig:numeric_sim}. The extracted $E_{\rm KW}$ is converted to vortex tilt angle by noting that
\begin{equation}
 \frac{\mathcal{E}_{\rm KW}}{\rho_{\rm s} L_{\rm tot}} = E_{\rm KW} \ell^2\,,
 \label{eq:EKW}
\end{equation}
and using Eqs.~\eqref{eq:Psi} and \eqref{eq:epsilonKW} to calculate $\Psi$, which in turn is converted to $\theta$ using Eq.~\eqref{eq:tantheta} or to $\theta_{\rm tot}$ using Eq.~\eqref{eq:totalAngle}. The simulations, Fig.~\ref{fig:numeric_sim}, reproduce most experimentally observed features. First, the initial buildup timescale for the turbulent state in the experiments coincides with the buildup time for $E_{\rm IW}$. Second, the necessity to drive the system for at least an hour to reach a steady state, Fig.~\ref{fig:relax_vs_modtime}, is in line with the buildup time for $E_{\rm KW}$. Third, the saturation level for the average vortex tilt angle agrees with the experimental value. Additionally, the simulations reproduce the initial relaxation phase during which vortex polarization remains at a constant level after stopping the modulations. Lastly, the simulations reproduce the exponential restoration of vortex polarization after the steady phase with similar time constant. The only unexplained observation, Fig.~\ref{fig:t00_vs_temp}, is the non-zero value of $t_{\rm g}(\Omega_0)$, which we believe to originate from features not included in our phenomenological model, such as excess energy stored in near-resonant inertial waves or geostrophic modes.

We also extract the steady state values of $E_{\rm KW}$ as a function of $\Omega_1$ after driving the system for 3 hours. The results for three different values of $\beta$ are shown in Fig.~\ref{fig:numeric_sim_saturation}. For all simulated values of $\beta \sim 1$, the simulations qualitatively reproduce the observed apparent saturation of $\lambda_{\rm eff}$ seen in the experiments, Fig.~\ref{fig:freq-satur}, where the signal is proportional to $f \propto -\sin^2\theta$.


\begin{thebibliography}{10}
\bibitem{SmoothSurfNote}
\bibinfo{note}{Here the upper spectrometer is located much closer to a surface
  than the bottom one, while the response is observed first in the bottom
  spectrometer. This suggests that the coupling between the quantized vortices
  and a smooth surface is negligible in comparison with that between the
  vortices and a rough surface. Even smoother surfaces may be produced in cold
  atom experiments, where a uniform trapping potential can be provided by
  repulsive laser light.}
\end{thebibliography}
\end{document}